\documentclass[journal]{IEEEtran}

\usepackage{chngpage}
\usepackage{amssymb}
\usepackage{color}
\usepackage{multirow}
\usepackage{mathrsfs}
\usepackage{xcolor}

\newtheorem{theorem}{Theorem}

\newtheorem{remark}{Remark}
\newtheorem{definition}{Definition}
\newtheorem{conjecture}{Conjecture}

\newcommand{\tagarray}{\mbox{}\refstepcounter{equation}$(\theequation)$}

\usepackage{cite}

\ifCLASSINFOpdf
  \usepackage[pdftex]{graphicx}
  \graphicspath{{../pdf/}{../jpeg/}}
  \DeclareGraphicsExtensions{.pdf,.jpeg,.png}
\else
  \usepackage[dvips]{graphicx}
  \graphicspath{{../eps/}}
  \DeclareGraphicsExtensions{.eps}
\fi
\usepackage[cmex10]{amsmath}
\usepackage{array}
\usepackage[tight,footnotesize]{subfigure}
\usepackage{microtype}
\usepackage{rotating}
\usepackage{epstopdf}
\begin{document}

\title{Optimal Coding Schemes for the Three-Receiver AWGN Broadcast Channel with Receiver Message Side Information}

\author{\IEEEauthorblockN{Behzad Asadi, Lawrence Ong, and Sarah J.\ Johnson}\thanks{The authors are with the School of Electrical Engineering and Computer Science, The University of Newcastle, Australia (e-mail: behzad.asadi@uon.edu.au, lawrence.ong@cantab.net, sarah.johnson@newcastle.edu.au). This work is supported by the Australian Research Council under grants FT110100195, FT140100219, and DP150100903.

This paper was presented in part at the 2014 IEEE International Symposium on Information Theory (ISIT 2014), and the 2014 IEEE Information Theory Workshop (ITW 2014).}
\IEEEauthorblockA{}}

\maketitle

\begin{abstract}
 This paper investigates the capacity region of the three-receiver AWGN broadcast channel where the receivers (i) have private-message requests, and (ii) may know some of the messages requested by other receivers as side information. We first classify all 64 possible side information configurations into eight groups, each consisting of eight members. We next construct transmission schemes, and derive new inner and outer bounds for the groups. This establishes the capacity region for 52 out of 64 possible side information configurations. For six groups (i.e., groups~1, 2, 3, 5, 6, and 8 in our terminology), we establish the capacity region for all their members, and show that it tightens both the best known inner and outer bounds. For group~4, our inner and outer bounds tighten the best known inner bound and/or outer bound for all the group members. Moreover, our bounds coincide at certain regions, which can be characterized by two thresholds. For group~7, our inner and outer bounds coincide for four members, thereby establishing the capacity region. For the remaining four members, our bounds tighten both the best known inner and outer bounds.
\end{abstract}


\IEEEpeerreviewmaketitle
\section{Introduction}
\IEEEPARstart{W}{e} consider \textit{private-message} broadcasting over the three-receiver additive white Gaussian noise (AWGN) broadcast channel where each receiver may know some of the messages requested by other receivers as side information. We investigate the capacity region of the channel for \textit{all possible side information configurations}.

\subsection{Background} 
Broadcast channels model communication networks where one transmitter wishes to transmit a number of messages to multiple receivers~\cite{BC}. The capacity region of broadcast channels is not known in general, except for a few special classes, e.g., degraded broadcast channels, which include AWGN broadcast channels~\cite{AWGNBCConverse}. 

In broadcast channels, the receivers may know some of the source messages a priori, referred to as receiver message side information. This is motivated by applications such as multimedia broadcasting with packet loss, and the downlink phase of multi-way relay channels~\cite{MWRCFullExchange}. The capacity region of broadcast channels with receiver message side information is known where each receiver needs to decode \textit{all} the source messages (or equivalently, all the source messages not known a priori)~\cite{SWoverBC,BCwithSI2UsersOechtering}.

However, the case where the receivers need not decode all the source messages remains unsolved to date. Kramer et al. established the capacity region of the two-receiver \textit{memoryless} broadcast channel where one of the receivers need not decode all the source messages~\cite{BCwithSI2UsersKramer}. The capacity region of the two-receiver \textit{AWGN} broadcast channel is known for all message requests and side information configurations~\cite{BCwithSI2UsersGeneral}. Extending the results to three or more receivers is ``highly nontrivial''~\cite{BCwithSI2UsersGeneral}. Oechtering et al.\ characterized the capacity region of the \textit{three-receiver} less noisy broadcast channel where (i) only two receivers possess side information and (ii) the request of the third receiver is only restricted to a common message demanded by all the receivers~\cite{BCwithSI3UsersCommonMessage}.

\begin{figure}[t]
	\centering
	\includegraphics[width=0.42\textwidth]{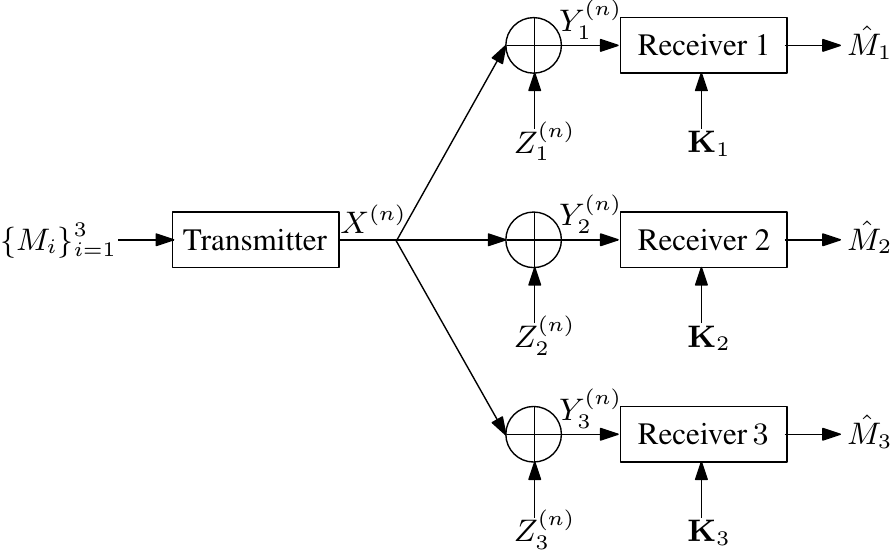}
	\caption{The three-receiver AWGN broadcast channel with receiver message side information, where $\{M_i\}_{i=1}^3$ are the source messages, each demanded by one receiver, and  $\mathbf{K}_i \subseteq \{M_1,M_2,M_3\}\setminus\{M_i\}$ is the set of messages known to receiver~$i$ a priori.} 
	\label{AWGNBCModelFig}
\end{figure}

\subsection{System Model and Problem Classification}\label{AWGN BC with SI}
This paper considers the three-receiver AWGN broadcast channel where the receivers have private-message requests. In the channel model under consideration, as depicted in Fig.~\ref{AWGNBCModelFig}, the channel-output sequence at receiver~$i$, $Y_{i}^{(n)}=\left(Y_{i,1},Y_{i,2},\ldots,Y_{i,n}\right),\;\,i=1,2,3$, is the sum of the transmitted codeword, $X^{(n)}=\left(X_{1},X_{2},\ldots,X_{n}\right)$, and an independent and identically distributed (i.i.d.) noise sequence, $Z_i^{(n)}=\left(Z_{i,1},Z_{i,2},\ldots,Z_{i,n}\right)$, with a normal distribution, $Z_i\sim \mathcal{N}\left(0, N_i\right)$. We represent random variables using upper-case letters, and their realizations using the corresponding lower-case letters. The input and the output alphabets of the channel are denoted by $\mathcal{X}$ and $\mathcal{Y}_i,\;\,i=1,2,3$, respectively. This channel is stochastically degraded, and without loss of generality, we can assume that receiver $1$ is the strongest and receiver $3$ is the weakest in the sense that $N_1\leq N_{2} \leq N_{3}$.

The transmitted codeword has a power constraint of $\sum_{j=1}^{n}E\left(X_j^2\right)\hspace{-3pt}\leq\hspace{-3pt}nP$ and is a function of source messages, $\{M_i\}_{i=1}^3$, which are independent. $M_i$ is intended for receiver~$i$, and is an $nR_i$-bit message, i.e., its transmission rate is $R_i$ bits per channel use. $M_i$ is uniformly distributed over the set $\mathcal{M}_i$ which is the set of all binary vectors of length $nR_i$, i.e., the cardinality of $\mathcal{M}_i$ is $2^{nR_i}$. We define the set $\mathbf{K}_i\subseteq \{M_1,M_2,M_3\}\setminus\{M_i\}$ as the set of messages known to receiver~$i$ a priori. 

A $\left(2^{nR_1},2^{nR_2},2^{nR_3},n\right)$ code for the channel consists of an encoding function
\begin{align*}
f: \mathcal{M}_1\times \mathcal{M}_2 \times \mathcal{M}_3 \rightarrow \mathcal{X}^{(n)},
\end{align*}
where $\times$ denotes the Cartesian product, and $\mathcal{X}^{(n)}$ denotes the $n$-fold Cartesian product of $\mathcal{X}$, i.e., a codeword, $x^{(n)}=f(m_1,m_2,m_3)$, is generated for each $(m_1,m_2,m_3)$. This code also consists of decoding functions
\begin{align*}
g_i: \mathcal{Y}_i^{(n)}\times \mathcal{K}_i\rightarrow \mathcal{M}_i,\;\;i=1,2,3,
\end{align*}
where, e.g., if $\mathbf{K}_1=\{M_2,M_3\}$, we have $\mathcal{K}_1=\mathcal{M}_2\times\mathcal{M}_3$. The decoded $M_i$ at receiver $i$ is $\hat{M}_i=g_i\left(Y_i^{(n)},\mathbf{K}_i\right)$, and the average probability of error for this code is defined as
\begin{align*}
P_e^{(n)}=P\left(\left(M_1,M_2,M_3\right)\neq(\hat{M}_1,\hat{M}_2,\hat{M}_3)\right).
\end{align*}

\begin{figure}[t]
	\centering
	\includegraphics[width=0.15\textwidth]{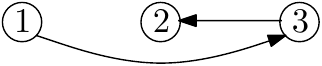}
	\caption{A sample side information graph where receiver 1 knows $M_3$, and receiver 3 knows $M_2$.}
	\label{Fig:SampleSIGraph}
\end{figure}

\begin{definition}
	A rate triple $(R_1,R_2,R_3)$ is said to be \textit{achievable} for the channel if there exists a sequence of $\left(2^{nR_1},2^{nR_2},2^{nR_3},n\right)$ codes with $P_e^{(n)}\rightarrow 0$ as $n \rightarrow \infty$.
\end{definition}
\begin{definition}
	The \textit{capacity region} of the channel is the closure of the set of all achievable rate triples $(R_1,R_2,R_3)$.
\end{definition}

The side information configuration of the channel is represented by a \textit{side information graph}, $\mathcal{G}=\left(\mathcal{V}_\mathcal{G},\mathcal{A}_\mathcal{G}\right)$, where $\mathcal{V}_\mathcal{G}=\{1,2,3\}$ is the set of \textit{vertices} and $\mathcal{A}_\mathcal{G}$ is the set of \textit{arcs}. An arc from vertex~$i$ to vertex~$j$, denoted by $(i\rightarrow j)$, exists if and only if receiver~$i$ knows $M_j$. The set of out-neighbors of vertex~$i$ is then $\mathcal{O}_i\triangleq\{j\mid (i\rightarrow j)\in\mathcal{A}_\mathcal{G}\}=\{j\mid M_j\in\mathbf{K}_i\}$. A sample side information graph is shown in Fig.~\ref{Fig:SampleSIGraph}.

\begin{figure}[t]
	\centering
	\includegraphics[width=0.35\textwidth]{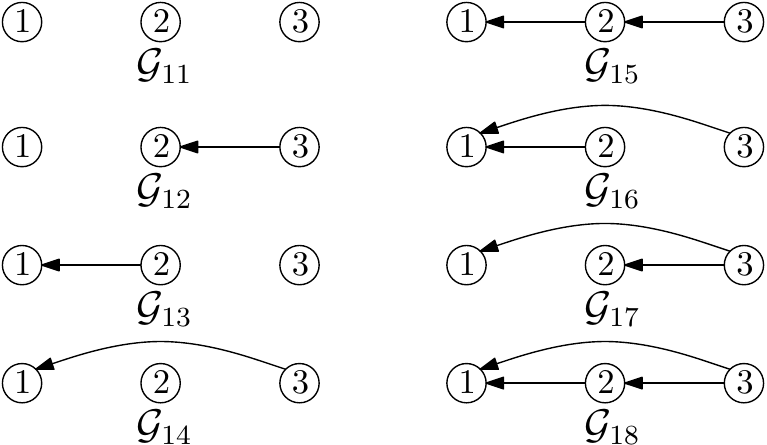}
	\caption{Graphs, $\{\mathcal{G}_{1i}\}_{i=1}^{8}$, capturing if each receiver knows the message(s) requested by stronger receiver(s).}
	\label{Fig:GroupLeaders}
\end{figure}

\begin{figure}[t]
	\centering
	\includegraphics[width=0.35\textwidth]{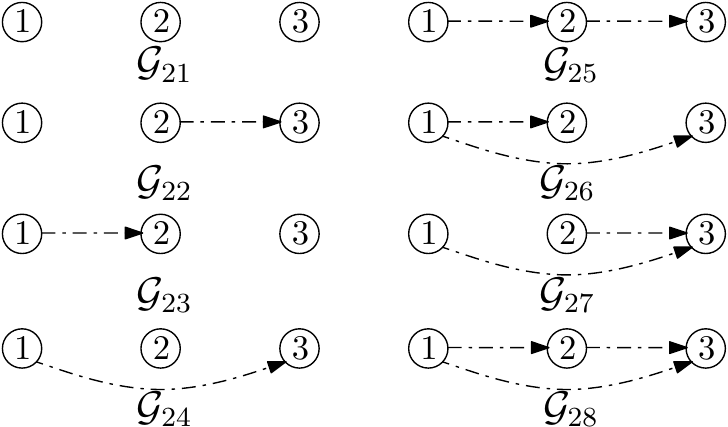}
	\caption{Graphs, $\{\mathcal{G}_{2j}\}_{j=1}^{8}$, capturing if each receiver knows the message(s) requested by weaker receiver(s).}
	\label{Fig:Subgraphs}
\end{figure}

We classify all 64 possible side information configurations into eight groups, each consisting of eight members. Any side information graph is the union of $\mathcal{G}_{1i}$, depicted in Fig. \ref{Fig:GroupLeaders}, and $\mathcal{G}_{2j}$, depicted in Fig. \ref{Fig:Subgraphs}, for some unique $i$ and $j$ where $i,j\in\{1,2,\ldots,8\}$. According to our proposed classification, the side information graphs $\{\mathcal{G}_{1i}\}_{i=1}^{8}$  are considered as the \textit{group leaders}, and group $i$ consists of the side information graphs formed from the union of the group leader, $\mathcal{G}_{1i}$, with each of the $\{\mathcal{G}_{2j}\}_{j=1}^8$. For instance, group 6 is the set $\{\mathcal{G}_{16}\cup\mathcal{G}_{2j}\}_{j=1}^8$.

\subsection{Existing Results and Contributions}
The capacity region of the channel is known where there is no side information at the receivers~\cite{AWGNBCConverse}, i.e., in our notation $\mathcal{G}_{11}\cup\mathcal{G}_{21}$. Oechtering et al.~\cite{OechteringG12G22G13G23} established the capacity region for two side information configurations, which in our notation are $\mathcal{G}_{12}\cup\mathcal{G}_{22}$ and $\mathcal{G}_{13}\cup\mathcal{G}_{23}$; they~\cite{OechteringG14G24} also derived the best known inner bound for $\mathcal{G}_{14}\cup\mathcal{G}_{24}$. These side information configurations correspond to the cases where two of the receivers know each other's requested messages and the third receiver has no side information.  

The best known inner and outer bounds for the remaining 60 side information configurations, and the best known outer bound for $\mathcal{G}_{14}\cup\mathcal{G}_{24}$ are within a constant gap of the capacity region~\cite{BCwithSI3UsersPrivateMessage}. The inner bound uses a separate index and channel coding scheme, developed based on the deterministic approach~\cite{Deterministic}. The outer bound is a polyhedron developed using Fano's inequality. 

\begin{figure}[t]
	\centering
	\includegraphics[width=0.45\textwidth]{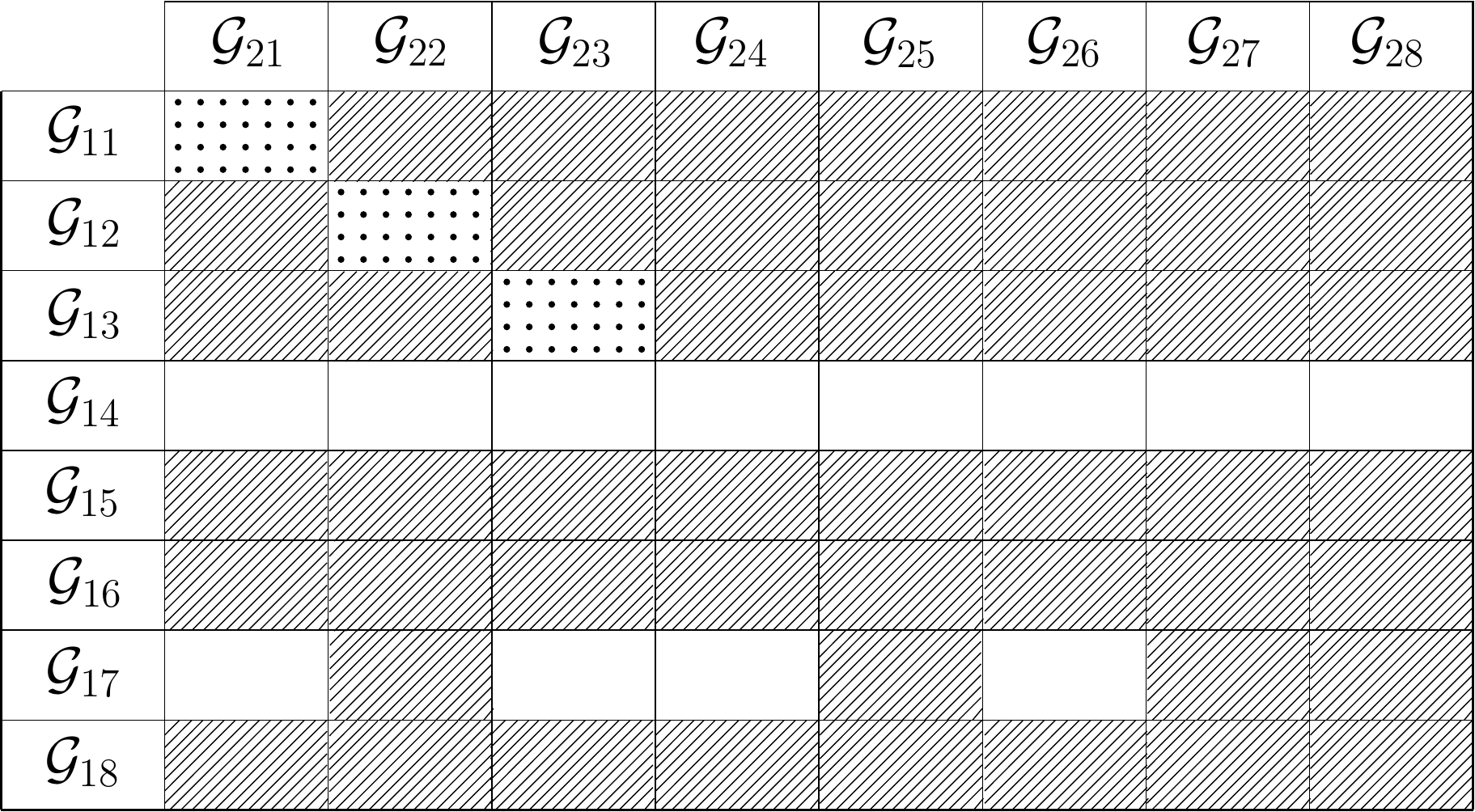}
	\caption{Result summary, where each cell represents the configuration formed from the union of $\mathcal{G}_{1i}$ and $\mathcal{G}_{2j}$ in its corresponding row and column (each row represents one group). The dotted cells show configurations whose capacity region was known prior to this work, the hatched cells show configurations whose capacity region is established in this work, and the blank cells show configurations with improved inner bound and/or outer bound but with capacity region remaining unknown.}
	\label{Fig:ResultSummary}
\end{figure}

We investigate the capacity region of the channel for all 64 possible side information configurations. One of the difficulties in deriving the capacity region for all side information configurations is to find a unified scheme. Considering the three-receiver AWGN broadcast channel without side information, i.e., $\mathcal{G}_{11}\cup\mathcal{G}_{21}$, it is well-known that the simple superposition of individual codewords, each carrying one message, can achieve the capacity region. For $\mathcal{G}_{11}\cup\mathcal{G}_{21}$, it can be easily verified that a stronger receiver is always able to decode the message of a weaker receiver if the weaker receiver is able to decode it; this means that knowing the message of a weaker receiver as side information does not help, and hence the same transmission scheme is optimal for all the members of group~1. Our proposed classification, developed based on this idea, facilitates the problem of having a unified scheme by grouping the side information configurations that lead to the same capacity-achieving transmission scheme for each group. We construct transmission schemes, and derive inner and outer bounds for different groups. For six groups, i.e., all the groups except groups~4 and 7, we establish the capacity region for all the group members. This result also shows the looseness of both the best known inner and outer bounds for these groups. For group~4, we improve the best known inner bound and/or outer bound for all the group members; our bounds coincide at certain regions which can be characterized by two thresholds. For group~7, we establish the capacity region for four members. For the remaining four members, we reduce the gap between the best known inner and outer bounds by improving both. Fig.~\ref{Fig:ResultSummary} shows a summary of our results.

In a concurrent work with the preliminary published versions \cite{Capacity3UsersPrivateMessage, Group4andGroup7} of this work, Sima et al.~\cite{Capacity3UsersPrivateMessageParallel} established the capacity region for 46 out of 64 possible side information configurations using different transmission schemes. These 46 configurations are a subset of the 52 side information configurations whose capacity region is established in our work.

\begin{table*}[t]
	\begin{footnotesize}
	\caption{Capacity region and proposed capacity-achieving transmission scheme for different groups}
	\vspace{-13pt}
	\begin{center}
		{\renewcommand{\arraystretch}{2}
			\begin{tabular}{|l|l|l|}
					\hline
					\hspace{-4pt}Group & Transmission Scheme & Capacity Region\\
					\hline\hline
					\hspace{-4pt}Group 1\hspace{-5pt} &\hspace{-4pt}$\color{blue}x_1^{(n)}\left(m_1\right)+x_2^{(n)}\left(m_2\right)+x_3^{(n)}\left(m_3\right)$
					&\hspace{-4pt}\color{blue}{$R_1 < C\left(\frac{\alpha_1P}{N_1}\right)$, $R_2<C\left(\frac{\alpha_2P}{\alpha_1P+N_2}\right)$, $R_3<C\left(\frac{\alpha_3P}{(\alpha_1+\alpha_2)P+N_3}\right)$}\\
					\hline
					\hspace{-4pt}Group 2\hspace{-5pt} &\hspace{-4pt}$\color{blue}x_1^{(n)}\left(m_1\right)+x_2^{(n)}\left([m_2,m_3]\right)$\hspace{-5pt}
					&\hspace{-4pt}\multirow{2}{*}{\color{blue}{$R_1 < C\left(\frac{\alpha_1P}{N_1}\right)$, $\sum_{i\in \{2,3\}\setminus\mathcal{O}_2}R_i<C\left(\frac{\alpha_2P}{\alpha_1P+N_2}\right)$, $R_3<C\left(\frac{\alpha_2P}{\alpha_1P+N_3}\right)$}}\\
					&\hspace{-5pt}$\mathcal{G}_{12} \cup \mathcal{G}_{22}$: $\color{blue}x_1^{(n)}\left(m_1\right)+x_2^{(n)}\left(m_2\oplus m_3\right)$&\\
					\hline
					\hspace{-4pt}Group 3\hspace{-5pt}  &$\hspace{-4pt}\color{blue}x_1^{(n)}\left([m_1,m_2]\right)+x_2^{(n)}\left(m_3\right)$
					&\hspace{-4pt}\color{blue}{${\sum_{i\in\{1,2\}\setminus\mathcal{O}_1}} R_i < C\left(\frac{\alpha_1 P}{N_1}\right)$, $R_2 < C\left(\frac{\alpha_1 P}{N_2}\right)$, $R_3 < C\left(\frac{\alpha_2 P}{\alpha_1 P + N_3}\right)$}\\
					\hline
					\hspace{-4pt}Group 5\hspace{-5pt} &\hspace{-4pt}$\color{blue}x_1^{(n)}\left([m_1,m_2]\right)+x_2^{(n)}\left([m_2,m_3]\right)$
					&\hspace{-4pt}\color{blue}{${\sum_{i\notin \mathcal{O}_1}} R_i < C\left(\frac{P}{N_1}\right)$, $R_1 < C\left(\frac{\alpha_1 P}{N_1}\right)$, ${\sum_{i\notin \mathcal{O}_2}} R_i  < C\left(\frac{P}{N_2}\right)$, $R_3<C\left(\frac{\alpha_2 P}{\alpha_1 P + N_3}\right)$}\\
					&\hspace{-5pt}$\mathcal{G}_{15} \cup \mathcal{G}_{22}$: \hspace{-4pt}$\color{blue}x_1^{(n)}\hspace{-3pt}\left([m_1,m_2\hspace{-1pt}\oplus\hspace{-1pt} m_3]\right)\hspace{-2pt}+\hspace{-2pt}x_2^{(n)}\hspace{-3pt}\left(m_2\hspace{-1pt}\oplus\hspace{-1pt} m_3\right)\hspace{-6pt}$ &\hspace{-5pt}$\mathcal{G}_{15} \cup \mathcal{G}_{22}$: \hspace{-4pt}\color{blue}{$R_1\hspace{-2pt}+\hspace{-2pt}\max\{R_2,R_3\}\hspace{-2pt}<\hspace{-2pt}C\hspace{-2pt}\left(\frac{P}{N_1}\right)$,\hspace{-2pt} $R_1\hspace{-2pt}<\hspace{-2pt}C\hspace{-2pt}\left(\frac{\alpha_1 P}{N_1}\right)$, \hspace{-4pt} $R_2\hspace{-2pt}<\hspace{-2pt}C\hspace{-2pt}\left(\frac{P}{N_2}\right)$, \hspace{-2pt}$R_3\hspace{-2pt}<\hspace{-2pt}C\hspace{-2pt}\left(\frac{\alpha_2P}{\alpha_1 P + N_3}\right)\hspace{-7pt}$}\\
					\hline
					\hspace{-4pt}Group 6\hspace{-5pt}  &\hspace{-4pt}$\color{blue}x_1^{(n)}\left([m_1,m_2]\right)+x_2^{(n)}\left([m_1,m_3]\right)$
					&\hspace{-4pt}\color{blue}{${\sum_{i\notin \mathcal{O}_1}} R_i < C\left(\frac{P}{N_1}\right)$, $R_2 < C\left(\frac{\alpha_1 P}{N_2}\right)$, $R_3 < C\left(\frac{\alpha_2 P}{\alpha_1 P + N_3}\right)$} \\ 
					\hline
					\hspace{-4pt}Group 8\hspace{-5pt}
					&\hspace{-4pt}$\color{blue}x^{(n)}\left([m_1,m_2,m_3]\right)$
					&\hspace{-4pt}\color{blue}{$\sum_{i\notin \mathcal{O}_1} R_i < C\left(\frac{P}{N_1}\right)$, $\sum_{i\notin \mathcal{O}_2} R_i < C\left(\frac{P}{N_2}\right)$, $R_3 < C\left(\frac{P}{N_3}\right)$}\\
					&\hspace{-5pt}$\mathcal{G}_{18} \cup \mathcal{G}_{22}$: $\color{blue}x^{(n)}\left([m_1,m_2\oplus m_3]\right)$ &\hspace{-5pt}$\mathcal{G}_{18} \cup \mathcal{G}_{22}$:\color{blue}{$\;R_1+\max \{R_2,R_3\} < C\left(\frac{P}{N_1}\right)$, $R_2 < C\left(\frac{P}{N_2}\right)$, $R_3 < C\left(\frac{P}{N_3}\right)$}\\
					\hline
			\end{tabular}}
			\label{TransmissionSchemes}
		\end{center}
	\end{footnotesize}
\end{table*}

\section{Capacity Region for Six Groups}
In this section, we first establish the capacity region for all the groups except groups 4 and 7, stated as Theorem \ref{maintheorem}. We then demonstrate the looseness of the best known inner and outer bounds~\cite{BCwithSI3UsersPrivateMessage} for these groups.

\subsection{Capacity Region for Groups 1, 2, 3, 5, 6, and 8}\label{section:capacity}

\begin{theorem}\label{maintheorem} The capacity region and the optimal transmission scheme for the three-receiver AWGN broadcast channel with private-message requests and side information configurations in groups~1, 2, 3, 5, 6 and~8 are shown in Table~\ref{TransmissionSchemes}. The capacity region for each configuration is the closure of the set of all rate triples $(R_1,R_2,R_3)$, each satisfying the conditions in the respective row for some $\alpha_\ell \geq 0$ where $\sum_{\text{all}\,\ell}\,\alpha_{\ell}=1$, and $C(t) \triangleq\frac{1}{2}\log(1+t)$.	
\end{theorem}

Before proving Theorem~\ref{maintheorem}, we summarize our proposed capacity-achieving transmission schemes shown in Table \ref{TransmissionSchemes}. All the members of each group use the same transmission scheme with the exception of one member in each of groups~2, 5 and~8.

Where the codebook of the transmission scheme is composed of multiple subcodebooks, the transmitted codeword, $x^{(n)}$, is constructed from the linear superposition of multiple codewords, $\sum_{\ell=1}^L x_\ell^{(n)}$ where $L$ is the number of subcodebooks. Each subcodebook consists of i.i.d. codewords, $x_\ell^{(n)}$, generated according to an independent normal distribution $X_\ell \sim \mathcal{N}(0,\alpha_\ell P)$, where $\alpha_\ell\hspace{-2pt}\geq\hspace{-2pt}0$ and $\sum_{\ell=1}^L \alpha_\ell=1$ to satisfy the transmission power constraint. Multiplexing coding~\cite{MultiplexedCoding} and index coding~\cite{Index Coding} are employed to construct the subcodebooks.

In multiplexing coding, two or more messages are first bijectively mapped to a single message, and then, the codewords are generated for this message. For instance, the first subcodebook of group~3 is constructed using multiplexing coding. In this scheme, a single message $M_\text{m}=[M_1,M_2]$ is first formed from $M_1$ and $M_2$, where square brackets, $[\cdot]$, denote a bijective map. Then, the codewords of the first subcodebook are generated for this single message, $M_\text{m}$, which is an $n(R_1+R_2)$-bit message. 

In index coding (which is also called network coding \cite{NetworkCoding} in some of the works on broadcast channels), the transmitter XORs the messages to accomplish compression prior to channel coding. 
The same function can also be achieved using modulo addition~\cite{BCwithSI2UsersKramer}. 
The transmission scheme for the exceptions in groups 2, 5 and 8  uses index coding. In these schemes, $M_\text{x}=M_2\oplus M_3$ is first formed, where $\oplus$ denotes the bitwise XOR with zero padding for messages of unequal length, i.e., $M_\text{x}$ is an $n\max\{R_2,R_3\}$-bit message. Then, the messages $M_1$ and $M_\text{x}$ are fed to the channel encoder (which performs multiplexing coding and superposition coding).

We now present the achievability proof of Theorem~\ref{maintheorem}, which elaborates on our proposed transmission schemes. We present the converse proof in the appendix.
\begin{IEEEproof}(\textit{Achievability})
The given rate region for each group is achieved using the transmission scheme given in Table~\ref{TransmissionSchemes}, and the following decoding scheme for the group. 

\textit{Group 1}: The decoders employ successive decoding where, at receiver $i$, $x_j^{(n)}$ is decoded while $\sum_{\ell<j}x_\ell^{(n)}$ is treated as noise starting from $j=3$ down to $j=i$. Then receiver~$i$ can reliably\footnote{We say that receiver~$i$ reliably decodes $M_j$ iff the probability of the decoded $M_j$ at this receiver being different from $M_j$, goes to zero as $n\rightarrow\infty$.} decode $M_i$ if $R_j<C\left(\frac{\alpha_j P}{\sum_{\ell=1}^{j-1}\alpha_\ell P+N_i}\right)\;\;\forall j\geq i$.

\textit{Group 2}: Receivers~2 and~3 decode $x_2^{(n)}$ while treating $x_1^{(n)}$ as noise. Receiver 2, depending on whether it knows $M_3$ or not, decodes $x_2^{(n)}$ over a set of $2^{nR_2}$ or $2^{n(R_2+R_3)}$ candidates, respectively. Then this receiver can reliably decode $M_2$ if $\sum_{i\in\{2,3\}\setminus\mathcal{O}_2} {R_i}< C(\frac{\alpha_2P}{\alpha_1P+N_2})$. Since receiver~3 knows $M_2$, it can reliably decode $M_3$ if $R_3< C(\frac{\alpha_2P}{\alpha_1P+N_3})$. Receiver~1 first decodes $x_2^{(n)}$ while treating $x_1^{(n)}$ as noise, and then decodes $x_1^{(n)}$.  This adds the condition $R_1<C(\frac{\alpha_1 P}{N_1})$ for achievability. For $\mathcal{G}_{12}\cup\mathcal{G}_{22}$, receiver~1 decodes $x_2^{(n)}$ over a set of $2^{n\max\{R_2,R_3\}}$ candidates, receiver~2 over a set of $2^{nR_2}$ candidates, and  receiver~3 over a set of $2^{nR_3}$ candidates. Note that the receivers use their side information during channel decoding. This is joint decoding, as opposed to separate decoding where the receivers do not use their side information during channel decoding. As an example, for $\mathcal{G}_{12}\cup\mathcal{G}_{21}$, we show the difference by writing the error events at receiver~3 for both joint decoding and separate decoding. We assume that the transmitted messages are equal to $\mathbf{0}$ by the symmetry of the code generation where $\mathbf{0}$ is the realization of the message with all bits equal to zero. Using separate decoding, receiver~3 decodes $\hat{m}_3$ if there exists a unique $\hat{m}_3$ such that $\left(X_2^{(n)}([m_2,\hat{m}_3]),Y_3^{(n)}\right)\in \mathcal{T}_\delta^{(n)}\text{ for some }m_2$, where $\mathcal{T}_\delta^{(n)}$ is the set of jointly $\delta$-typical $n$-sequences with respect to the considered distribution~\cite[p.\ 521]{ITBook}; otherwise an error is declared. Then the error events at receiver~3 are
\begin{align*}
&\mathcal{E}_{31}:\left(X_2^{(n)}([\mathbf{0},\mathbf{0}]),Y_3^{(n)}\right)\notin \mathcal{T}_\delta^{(n)},\\
&\mathcal{E}_{32}:\left(X_2^{(n)}([m_2,m_3]),Y_3^{(n)}\right)\in \mathcal{T}_\delta^{(n)}\hskip3pt\text{for some }m_3\neq\mathbf{0},m_2.
\end{align*}
 According to these error events, and the properties of joint typicality~\cite[Theorems 15.2.1 and 15.2.3]{ITBook}, receiver 3 can reliably decode $M_3$ if
\begin{align*}
R_2+R_3<C\left(\frac{\alpha_2P}{\alpha_1P+N_3}\right).
\end{align*}
However, using joint decoding, receiver~3 utilizes its side information, $m_2=\mathbf{0}$, during channel decoding, and decodes $\hat{m}_3$ if there exists a unique $\hat{m}_3$ such that $\left(X_2^{(n)}([\mathbf{0},\hat{m}_3]),Y_3^{(n)}\right)\in \mathcal{T}_\delta^{(n)}$; otherwise an error is declared. Then the error events at receiver~3 are
\begin{align*}
	&\mathcal{E}_{31}:\left(X_2^{(n)}([\mathbf{0},\mathbf{0}]),Y_3^{(n)}\right)\notin \mathcal{T}_\delta^{(n)},\\
	&\mathcal{E}_{32}:\left(X_2^{(n)}([\mathbf{0},m_3]),Y_3^{(n)}\right)\in \mathcal{T}_\delta^{(n)}\hskip10pt\text{for some }m_3\neq\mathbf{0},
\end{align*}
which show that receiver~3 can reliably decode $M_3$ if 
\begin{align*}
R_3<C\left(\frac{\alpha_2P}{\alpha_1P+N_3}\right).
\end{align*}
Hence, using separate decoding, we have the more restrictive condition $R_2+R_3<C(\frac{\alpha_2P}{\alpha_1P+N_3})$ which leads to a strictly smaller achievable rate region when $N_2<N_3$.

\textit{Group 3}: Receivers~1 and 2 first decode $x_2^{(n)}$ while treating $x_1^{(n)}$ as noise, and then decode $x_1^{(n)}$ using their side information. Receiver~3 only decodes $x_2^{(n)}$ while treating $x_1^{(n)}$ as noise. Hence, $\sum_{i\in\{1,2\}\setminus\mathcal{O}_1} {R_i}<C\left(\frac{\alpha_1 P}{N_1}\right)$, $R_2<C\left(\frac{\alpha_1 P}{N_2}\right)$, and $R_3< C(\frac{\alpha_2P}{\alpha_1P+N_3})$ are the sufficient conditions for achievability. 

\textit{Group 5} (all the members except $\mathcal{G}_{15}\cup\mathcal{G}_{22}$): \textit{Simultaneous decoding} is utilized for this group at receivers~1 and~2, and successive decoding at receiver~3. Receiver~3 decodes $x_2^{(n)}$ while treating $x_1^{(n)}$ as noise. Since receiver~3 knows $M_2$, it can reliably decode $M_3$ if $R_3< C(\frac{\alpha_2P}{\alpha_1P+N_3})$. For $\mathcal{G}_{15} \cup \mathcal{G}_{21}$, receiver~1 decodes $\hat{m}_1$ if there exists a unique $\hat{m}_1$ such that $\left(X_1^{(n)}([\hat{m}_1,m_2]),X_2^{(n)}([m_2,m_3]),Y_1^{(n)}\right)\in \mathcal{T}_\delta^{(n)}$ for some $m_2\text{ and }m_3$; otherwise an error is declared. For this member, assuming the transmitted messages are equal to $\mathbf{0}$, the error events at receiver~1 are
\begin{align*}
&\mathcal{E}_{11}:\left(X_1^{(n)}([\mathbf{0},\mathbf{0}]),X_2^{(n)}([\mathbf{0},\mathbf{0}]),Y_1^{(n)}\right)\notin \mathcal{T}_\delta^{(n)},\\
&\mathcal{E}_{12}:\left(X_1^{(n)}([m_1,\mathbf{0}]),X_2^{(n)}([\mathbf{0},\mathbf{0}]),Y_1^{(n)}\right)\in \mathcal{T}_\delta^{(n)}\\
&\hskip175pt\text{for some }m_1\neq\mathbf{0},\\
&\mathcal{E}_{13}:\left(X_1^{(n)}([m_1,\mathbf{0}]),X_2^{(n)}([\mathbf{0},m_3]),Y_1^{(n)}\right)\in \mathcal{T}_\delta^{(n)}\\
&\hskip140pt\text{for some }m_1\neq\mathbf{0}, m_3\neq\mathbf{0},\\
&\mathcal{E}_{14}:\left(X_1^{(n)}([m_1,m_2]),X_2^{(n)}([m_2,m_3]),Y_1^{(n)}\right)\in \mathcal{T}_\delta^{(n)}\\
&\hskip122pt\text{for some }m_1\neq\mathbf{0}, m_2\neq\mathbf{0}, m_3.
\end{align*}
Then receiver~1 can reliably decode $M_1$ if $R_1+R_2+R_3<C(\frac{P}{N_1})$ and $R_1<C(\frac{\alpha_1P}{N_1})$. For the other members, receiver~1 makes its decoding decision based on its extra side information. For example, if receiver~1 knows $M_3$, the mentioned conditions reduce to $R_1+R_2<C(\frac{P}{N_1})$ and $R_1<C(\frac{\alpha_1P}{N_1})$. Hence, $\sum_{i\notin \mathcal{O}_1}R_i<C(\frac{P}{N_1})$ and $R_1<C(\frac{\alpha_1P}{N_1})$ guarantee that receiver~1 can decode $M_1$ reliably.

Receiver~2 when $M_3$ is unknown a priori decodes $\hat{m}_2$ if there exists a unique $\hat{m}_2$ such that $\left(X_1^{(n)}([\mathbf{0},\hat{m}_2]),X_2^{(n)}([\hat{m}_2,m_3]),Y_2^{(n)}\right)\in \mathcal{T}_\delta^{(n)}$ for some $m_3$; otherwise an error is declared. For these members, the error events at receiver~2 are (receiver~2 knows $M_1$ a priori)
\begin{align*}
&\mathcal{E}_{21}: \left(X_1^{(n)}([\mathbf{0},\mathbf{0}]),X_2^{(n)}([\mathbf{0},\mathbf{0}]),Y_2^{(n)}\right)\notin \mathcal{T}_\delta^{(n)},\\
&\mathcal{E}_{22}:\left(X_1^{(n)}([\mathbf{0},m_2]),X_2^{(n)}([m_2,m_3]),Y_2^{(n)}\right)\in \mathcal{T}_\delta^{(n)}\\
&\hskip155pt\text{for some }\;m_2\neq\mathbf{0},m_3,
\end{align*}
which show that receiver~2 can reliably decode $M_2$ if $R_2+R_3<C(\frac{P}{N_2})$. If receiver~2 knows $M_3$ a priori, this condition reduces to $R_2<C(\frac{P}{N_2})$.

$\mathcal{G}_{15}\cup\mathcal{G}_{22}$: For this side information configuration, we use the same decoding scheme as for the other members of this group. Since receiver~3 knows $M_2$, it can reliably decode $M_3$ if $R_3< C(\frac{\alpha_2P}{\alpha_1P+N_3})$. Assuming the transmitted messages are equal to $\mathbf{0}$, which also yields $m_\text{x}=m_2\oplus m_3=\mathbf{0}$, the error events at receiver~1 are
\begin{align*}
&\mathcal{E}_{11}:\left(X_1^{(n)}([\mathbf{0},\mathbf{0}]),X_2^{(n)}(\mathbf{0}),Y_1^{(n)}\right)\notin \mathcal{T}_\delta^{(n)},\\
&\mathcal{E}_{12}:\left(X_1^{(n)}([m_1,\mathbf{0}]),X_2^{(n)}(\mathbf{0}),Y_1^{(n)}\right)\in \mathcal{T}_\delta^{(n)}\\
&\hskip174pt\text{for some }m_1\neq\mathbf{0},\\
&\mathcal{E}_{13}:\left(X_1^{(n)}([m_1,m_\text{x}]),X_2^{(n)}(m_\text{x}),Y_1^{(n)}\right)\in \mathcal{T}_\delta^{(n)}\\
&\hskip138pt\text{for some }m_1\neq\mathbf{0}, m_\text{x}\neq\mathbf{0},
\end{align*}
and the error events at receiver~2 are (receiver 2 knows $M_1$ and $M_3$ a priori)
\begin{align*}
&\mathcal{E}_{21}:\left(X_1^{(n)}([\mathbf{0},\mathbf{0}]),X_2^{(n)}(\mathbf{0}),Y_2^{(n)}\right)\notin \mathcal{T}_\delta^{(n)},\\
&\mathcal{E}_{22}:\left(X_1^{(n)}([\mathbf{0},m_2\oplus \mathbf{0}]),X_2^{(n)}(m_2\oplus \mathbf{0}),Y_2^{(n)}\right)\in \mathcal{T}_\delta^{(n)}\\
&\hskip173pt \text{for some }m_2\neq\mathbf{0}.
\end{align*}
These error events show that receiver~1 can reliably decode $M_1$ if $R_1+\max\{R_2,R_3\}<C(\frac{P}{N_1})$ and $R_1<C(\frac{\alpha_1 P}{N_1})$, and receiver~2 can reliably decode $M_2$ if $R_2<C(\frac{P}{N_2})$. 

\textit{Group 6}: Simultaneous decoding is utilized for this group at receiver~1, and successive decoding at receivers~2 and~3. At receiver~3, $x_2^{(n)}$ is decoded while $x_1^{(n)}$ is treated as noise. Since receiver~3 knows $M_1$, it can reliably decode $M_3$ if $R_3< C(\frac{\alpha_2P}{\alpha_1P+N_3})$. At receiver~2, $x_2^{(n)}$ is first decoded while $x_1^{(n)}$ is treated as noise, and then $x_1^{(n)}$ is decoded. Since receiver~2 knows $M_1$, it can reliably decode $M_2$ if $R_3< C(\frac{\alpha_2P}{\alpha_1P+N_3})$ and $R_2< C(\frac{\alpha_1P}{N_2})$. For $\mathcal{G}_{16} \cup \mathcal{G}_{21}$, receiver~1 decodes $\hat{m}_1$ if there exists a unique $\hat{m}_1$ such that $\left(X_1^{(n)}([\hat{m}_1,m_2]),X_2^{(n)}([\hat{m}_1,m_3]),Y_1^{(n)}\right)\in \mathcal{T}_\delta^{(n)}$ for some $m_2\text{ and }m_3$; otherwise an error is declared. For this member, assuming the transmitted messages are equal to $\mathbf{0}$, the error events at receiver~1 are
\begin{align*}
&\mathcal{E}_{11}:\left(X_1^{(n)}([\mathbf{0},\mathbf{0}]),X_2^{(n)}([\mathbf{0},\mathbf{0}]),Y_1^{(n)}\right)\notin \mathcal{T}_\delta^{(n)},\\
&\mathcal{E}_{12}:\left(X_1^{(n)}([m_1,m_2]),X_2^{(n)}([m_1,m_3]),Y_1^{(n)}\right)\in \mathcal{T}_\delta^{(n)}\\
&\hskip135pt\text{for some }m_1\neq\mathbf{0},m_2,m_3.
\end{align*}
Then receiver~1 can reliably decode $M_1$ if $R_1+R_2+R_3<C(\frac{P}{N_1})$. For the other members, receiver~1 makes its decoding decision based on its extra side information. For example, if receiver~1 knows $M_3$, $R_1+R_2+R_3<C(\frac{P}{N_1})$ reduces to $R_1+R_2<C(\frac{P}{N_1})$. Hence, $\sum_{i\notin \mathcal{O}_1}R_i<C(\frac{P}{N_1})$ guarantees that receiver~1 can reliably decode $M_1$.

\textit{Group 8}: In this group, each receiver decodes the correct $x^{(n)}$ over a set of valid candidates based on its side information. For instance, since receiver 3 knows $M_1$ and $M_2$ as side information, it can reliably decode $M_3$ if $R_3<C(\frac{P}{N_3})$.
\end{IEEEproof}

\subsection{On the Looseness of Prior Bounds} \label{section:looseness}
In this subsection, we demonstrate the looseness of the best known inner and outer bounds for the six groups in Section~\ref{section:capacity}.

The best known inner bound, which is achieved by a separate index and channel coding scheme (developed based on the deterministic approach), is the set of all rate triples $(R_1,R_2,R_3)$, each satisfying~\cite{BCwithSI3UsersPrivateMessage}
\begin{equation}\label{deterministicregion}
\sum_{i\in\mathcal{V}_\mathcal{S}}{R_i} < \underset{i\in\mathcal{V}_\mathcal{S}}{\max} A_i,
\vspace{-3pt}
\end{equation}
for all induced acyclic subgraphs, $\mathcal{S}$, of the side information graph, where $\mathcal{V}_\mathcal{S}$ is the vertex set of $\mathcal{S}$. In \eqref{deterministicregion}, $A_i=\sum_{\ell=i}^{3}{B_\ell},\;\;i=1,2,3,$ where $B_1=C(\frac{\alpha_1P}{N_1})$, $B_2=C(\frac{\alpha_2P}{\alpha_1 P+N_2})$, and $B_3=C(\frac{\alpha_3P}{(\alpha_1+\alpha_2)P+N_3})$ for some $\alpha_\ell\geq0$ such that $\sum_{\ell=1}^{3}{\alpha_\ell}=1$.

As an example, this region for $\mathcal{G}_{12} \cup \mathcal{G}_{21}$ can be achieved using the transmission scheme
\begin{align*}
x_1^{(n)}(m_{11})+x_2^{(n)}([m_{12},m_{21}])+x_3^{(n)}([m_{13},m_{22},m_3]),
\vspace{-1pt}
\end{align*}
and a separate decoding scheme (where side information is \textit{not utilized} during channel decoding). This scheme uses rate splitting where the message $M_1$ is divided into independent messages $\{M_{1i}\}_{i=1}^3$ at rates $\{R_{1i}\}_{i=1}^3$ such that  $R_1=\sum_{i=1}^{3}{R_{1i}}$; the message $M_2$ is also divided into independent messages $\{M_{2i}\}_{i=1}^2$ at rates $\{R_{2i}\}_{i=1}^2$ such that  $R_2=R_{21}+R_{22}$.
For $\mathcal{G}_{12} \cup \mathcal{G}_{21}$, we can verify the achievability of the region in \eqref{deterministicregion} using Fourier-Motzkin elimination subsequent to successive decoding where, at receiver $i$, $x_j^{(n)}$ is decoded while $\sum_{\ell<j}x_\ell^{(n)}$ is treated as noise starting from $j=3$ down to $j=i$.

\begin{figure}[t]
	\hskip-17pt\includegraphics[width=0.56\textwidth]{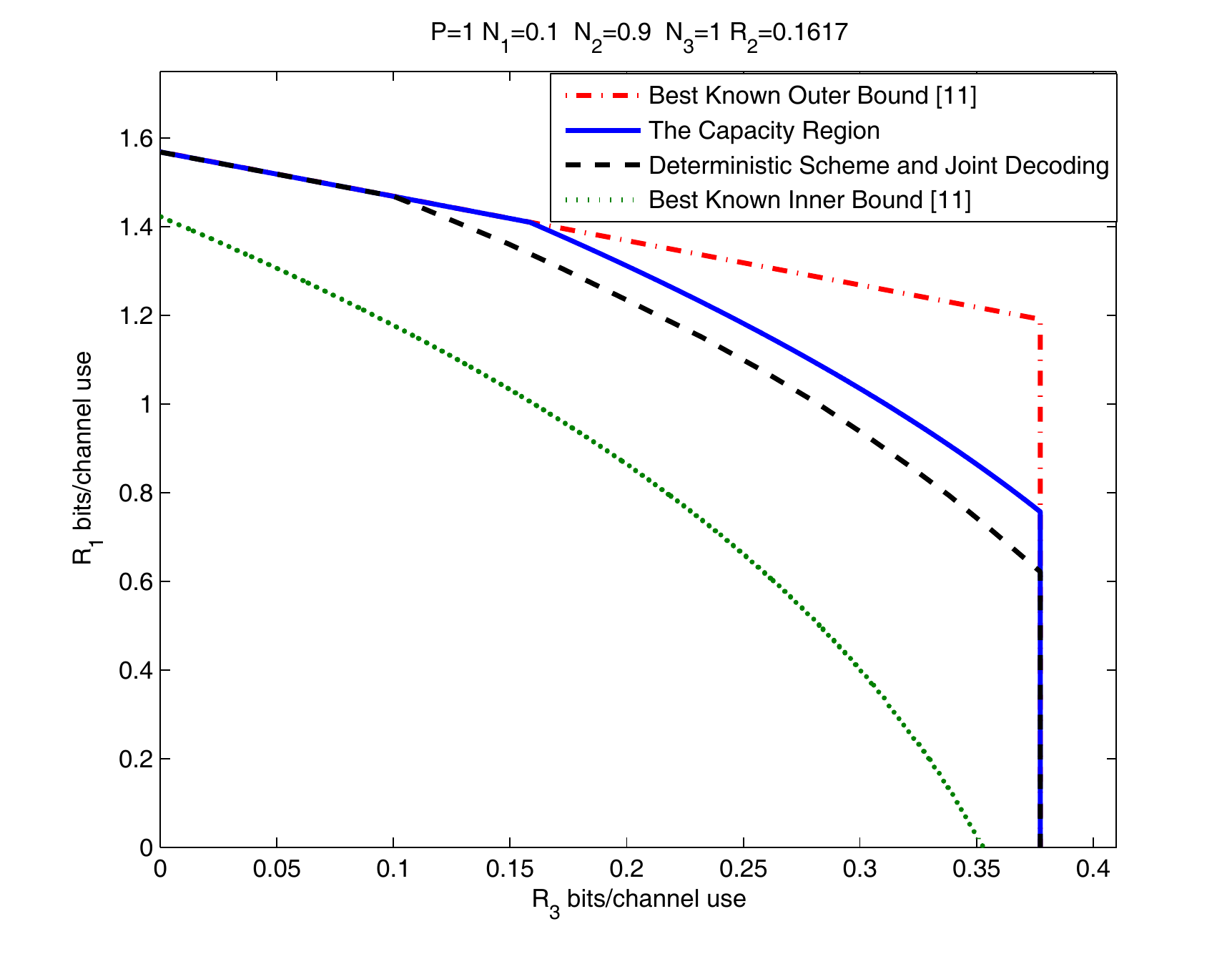}
	\caption{Capacity region, inner bound and outer bound comparison for $\mathcal{G}_{15}\cup\mathcal{G}_{21}$.} 
	\label{Comparison}
\end{figure}

We now show that the achievable rate region can be improved using the same encoding scheme, but \textit{utilizing} side information during successive decoding (i.e., joint decoding). Consider the given example ($\mathcal{G}_{12} \cup \mathcal{G}_{21}$). Using separate decoding, the receivers can reliably decode $x_3^{(n)}$ while $x_1^{(n)}+x_2^{(n)}$ is treated as noise if $R_{13}+R_{22}+R_3 <B_3$. Using joint decoding, we can relax this condition  to $R_{13}+R_3<B_3$ and $R_{13}+R_{22}+R_3 <B'_3$ where $B'_3 = C(\frac{\alpha_3P}{(\alpha_1+\alpha_2)P+N_2})$ ($B'_3\geq B_3$ for any choice of $\{\alpha_\ell\}_{\ell=1}^3$). This gives an improved achievable rate region for $\mathcal{G}_{12} \cup \mathcal{G}_{21}$.

This joint decoding approach can be used to strictly enlarge the rate region in \eqref{deterministicregion} for all the side information configurations in groups 2 to 8.\footnote{The capacity region for group~1 is the same as the capacity region of the three-receiver AWGN broadcast channel without receiver message side information.} Having demonstrated the suboptimality of the best known inner bound in~\eqref{deterministicregion}, it follows that our optimal inner bound must be strictly larger. For example, Fig. \ref{Comparison} depicts that the best known inner bound is strictly loose for $\mathcal{G}_{15}\cup\mathcal{G}_{21}$. This figure also shows that the encoding scheme developed based on the deterministic approach cannot achieve the capacity region even using the proposed joint decoding approach.

Here, we demonstrate the looseness of the best known outer bound for the six groups in Secion~\ref{section:capacity}. The best known outer bound states that if a rate triple $(R_1,R_2,R_3)$ is achievable, it must satisfy~\cite{BCwithSI3UsersPrivateMessage}
\begin{equation}\label{bestpriorouterbound}
\sum_{i\in\mathcal{V}_\mathcal{S}}{R_i} \leq \underset{i\in\mathcal{V}_\mathcal{S}}{\max}\;C\left(\frac{P}{N_i}\right),
\vspace{-0pt}
\end{equation}
for all induced acyclic subgraphs, $\mathcal{S}$, of the side information graph. This outer bound is a polyhedron, and is strictly loose for the six groups whose capacity region was established in Section~\ref{section:capacity} except group 8.\footnote{The capacity region for group 8 is a polyhedron.} This is because the capacity region for these groups is a function of $\alpha_\ell$, and therefore the capacity region has some curved surfaces. For example, Fig.~\ref{Comparison} depicts the looseness of this outer bound for $\mathcal{G}_{15}\cup\mathcal{G}_{21}$.

\section{Group 4: Inner and Outer Bounds}\label{Section:Group4}
In this section, we first derive inner and outer bounds for group 4. We then compare them with the best known ones, and characterize the regions where our bounds coincide. \textit{Dirty paper coding} \cite{DPC} proves to be useful for this group. Dirty paper coding is used when the channel between a transmitter and a receiver is affected by an interference $s^{(n)}$ which is known non-causally at the transmitter; codewords using this coding technique are denoted by $x_\ell^{(n)}(m,s^{(n)})$ which are functions of both the transmitted message $m$ and the interference $s^{(n)}$. We also employ the notion of \textit{enhanced channel} \cite{MIMOBC} for this group to tighten the best known outer bound. 
\subsection{Inner Bounds}
\begin{theorem}\label{theorem:innergroup4}
	A rate triple $\left(R_1,R_2,R_3\right)$ for the members $\mathcal{G}_{14}\cup\mathcal{G}_{2i},\;\,i=1,3,4,6,7,8$, is achievable if it satisfies
	\begin{align}
		\underset{i\in\{1,3\}\setminus\mathcal{O}_1}{\sum}\hskip-5ptR_i&\hskip-2pt<\hskip-2ptC\hskip-2pt\left(\hskip-2pt\frac{\alpha (1-\beta) P}{\alpha\beta P+(1-\alpha)P+N_1}\hskip-2pt\right)\hskip-3pt+\hskip-2ptC\hskip-2pt\left(\hskip-2pt\frac{\alpha \beta P}{N_1}\hskip-2pt\right),\label{inner41}\\
		R_2&\hskip-2pt<\hskip-2ptC\left(\frac{(1-\alpha)P}{\alpha\beta P+N_2}\right),\label{inner42}\\
		R_3&\hskip-2pt<\hskip-2ptC\hskip-2pt\left(\hskip-2pt\frac{\alpha (1-\beta) P}{\alpha\beta P+(1-\alpha)P+N_3}\hskip-2pt\right)\hskip-3pt+\hskip-2ptC\hskip-2pt\left(\hskip-2pt\frac{\alpha \beta P}{N_3}\hskip-2pt\right),\label{inner43}
	\end{align}
	for some $0\leq\alpha\leq1$ and $0\leq\beta\leq1$. For the members $\mathcal{G}_{14}\cup\mathcal{G}_{2i},\;\,i=2,5$, it is achievable if it satisfies
	\begin{align}
		R_1&<C\left(\frac{\alpha (1-\beta) P}{\alpha\beta P+(1-\alpha)P+N_1}\right)\hskip-3pt+\hskip-2ptC\left(\frac{\alpha \beta P}{N_1}\right),\label{inner44}\\
		\underset{i\notin\mathcal{O}_1}{\sum}R_i &<C\left(\frac{P}{N_1}\right),\label{inner45}\\
		R_2&<C\left(\hspace{-1pt}\frac{(1-\alpha)P}{\alpha\beta P+N_2}\right),\label{inner46}\\
		R_3&<C\left(\frac{\alpha (1-\beta) P}{\alpha\beta P+(1-\alpha)P+N_3}\right)\hskip-3pt+\hskip-2ptC\left(\frac{\alpha \beta P}{N_3}\right),\label{inner47}
	\end{align}
	for some $0\leq\alpha\leq1$ and $0\leq\beta\leq1$.
\end{theorem}

Before proving Theorem~\ref{theorem:innergroup4}, we explain the transmission schemes used to achieve the inner bounds. The inner bound for the members $\mathcal{G}_{14}\cup\mathcal{G}_{2i},\;\,i=1,3,4,6,7,8$, is achieved using the transmission scheme proposed by Oechtering et al.~\cite{OechteringG14G24}; the transmission scheme takes the form 
\begin{align}\label{scheme1group4}
x_1^{(n)}\hskip-2pt\left([m'_{1},m'_{3}],x_2^{(n)}\right)\hskip-3pt+\hskip-3ptx_2^{(n)}\hskip-2pt\left(m_2,x_3^{(n)}\right)\hskip-3pt+\hskip-3ptx_3^{(n)}\left([m''_{1},m''_{3}]\right).
\end{align}
The inner bound for the remaining two members, $\mathcal{G}_{14}\cup\mathcal{G}_{2i},\;\,i=2,5$, is achieved using the following proposed transmission scheme
\begin{multline}\label{scheme2group4}
x_1^{(n)}\left([m'_{1}\oplus m'_{31},m'_{32}],x_2^{(n)}\right)+\\ x_2^{(n)}\left([m_2,m_{31}],x_3^{(n)}\right)+x_3^{(n)}\left([m''_{1}\oplus m''_{31},m''_{32}]\right).
\end{multline}
The codebooks of both transmission schemes in \eqref{scheme1group4} and \eqref{scheme2group4} are formed from the linear superposition of three subcodebooks, where the first two subcodebooks are constructed employing dirty paper coding. In \eqref{scheme1group4} and \eqref{scheme2group4}, the third subcodebook consists of i.i.d.\ codewords $x_3^{(n)}$ generated according to $X_3\hspace{-2pt}\sim\mathcal{N}\left(0,\alpha(1-\beta)P\right)$ where $0\leq\alpha\leq1$ and $0\leq\beta\leq1$. By treating $x_3^{(n)}$ as interference for receiver 2 (known non-causally at the transmitter), the second subcodebook is constructed using dirty paper coding with the auxiliary random variable $U_2=X_2+\lambda_2 X_3$ where $X_2\sim\mathcal{N}\left(0,(1-\alpha)P\right)$, and $\lambda_2=\frac{(1-\alpha)P}{(1-\alpha)P+\alpha\beta P+N_2}$. Also, by treating $x_2^{(n)}$ as interference for receiver 3, the first subcodebook is constructed using dirty paper coding with the auxiliary random variable $U_1=X_1+\lambda_1 X_2$ where $X_1\sim\mathcal{N}\left(0,\alpha\beta P\right)$, and $\lambda_1=\frac{\alpha\beta P}{\alpha\beta P+N_3}$. $X_1$, $X_2$ and $X_3$ are also mutually independent.

Rate splitting is used in both transmission schemes. In \eqref{scheme1group4}, using rate splitting, the message $M_i,\;\;i=1,3,$ is divided into two independent messages $M'_{i}$ at rate $R'_{i}$, and $M''_{i}$ at rate $R''_{i}$ such that  $R_i=R'_{i}+R''_{i}$. In \eqref{scheme2group4}, $M_1$ is similarly divided into two independent messages $M'_{1}$ at rate $R'_{1}$, and $M''_{1}$ at rate $R''_{1}$ such that $R_1=R'_{1}+R''_{1}$. But $M_3$ is first divided into two independent messages $\{M_{3i}\}_{i=1}^2$ at rates $\{R_{3i}\}_{i=1}^2$ such that $R_3=R_{31}+R_{32}$. Each of $\{M_{3i}\}_{i=1}^2$ is then divided into two independent messages $M'_{3i}$ at rate $R'_{3i}$ and $M''_{3i}$ at rate $R''_{3i}$ such that $R_{3i}=R'_{3i}+R''_{3i}$.

We now present the proof of Theorem~\ref{theorem:innergroup4}.

\begin{IEEEproof}
The achievability of the inner bound in \eqref{inner41}--\eqref{inner43} is proved by using the transmission scheme in \eqref{scheme1group4}, and the following decoding methods. 

Receiver 3 first decodes $x_3^{(n)}$ while treating $x_1^{(n)}+x_2^{(n)}$ as noise. Since this receiver knows $M_1$, it can reliably decode $x_3^{(n)}$ if
\begin{align*}
R''_{3}<C\left(\frac{\alpha (1-\beta) P}{\alpha\beta P+(1-\alpha)P+N_3}\right).
\end{align*}
Receiver~3 then decodes $M'_{3}$ without being affected by $x_2^{(n)}$ due to dirty paper coding. After correctly decoding $x_3^{(n)}$, this receiver can reliably decode $M'_{3}$ if
\begin{align*}
R'_{3}<C\left(\frac{\alpha \beta P}{N_3}\right).
\end{align*}
Receiver 2 decodes $M_2$ while treating $x_1^{(n)}$ as noise. This receiver can reliably decode $M_2$ (without being affected by $x_3^{(n)}$ due to dirty paper coding) if
\begin{align}\label{group4condition2}
R_2<C\left(\frac{(1-\alpha)P}{\alpha\beta P+N_2}\right).
\end{align}
Receiver 1 first decodes $x_3^{(n)}$ while treating $x_1^{(n)}+x_2^{(n)}$ as noise, and then decodes $x_2^{(n)}$ while treating $x_1^{(n)}$ as noise. Hence, this receiver can reliably decode $x_2^{(n)}$ and $x_3^{(n)}$ if
\begin{align*}
\underset{i\in\{1,3\}\setminus\mathcal{O}_1}{\sum}R''_{i} &<C\left(\frac{\alpha (1-\beta) P}{\alpha\beta P+(1-\alpha)P+N_1}\right),
\end{align*}
and
\begin{align} \label{group4redundantcondition2}
R_2&<C\left(\frac{(1-\alpha)P}{\alpha\beta P+N_1}\right).
\end{align}
Considering \eqref{group4condition2}, \eqref{group4redundantcondition2} is redundant. Note that dirty paper coding is utilized for constructing the second subcodebook. However, when interference in dirty paper coding (here $x_3^{(n)}$) is known at both the transmitter and the receiver, the achievability condition is not a function of $\lambda_2$. This can be confirmed by
\begin{align}
R_2&<I(U_2;Y_1,X_3)-I(U_2;X_3)=I(U_2;Y_1\mid X_3)\nonumber\\
&=I(X_2+\lambda_2 X_3;X_1+X_2+X_3+Z_1\mid X_3)\nonumber\\
&=I(X_2;X_1+X_2+Z_1),\label{DPCknownReceiverinterference}
\end{align}
which leads to the condition in \eqref{group4redundantcondition2}.
After decoding $x_2^{(n)}$ and $x_3^{(n)}$, receiver~1 decodes $x_1^{(n)}$. This adds the condition
\begin{align*}
\underset{i\in\{1,3\}\setminus\mathcal{O}_1}{\sum}R'_{i} &<C\left(\frac{\alpha \beta P}{N_1}\right),
\end{align*}
for achievability, where an argument similar to \eqref{DPCknownReceiverinterference} is required. 
Fourier-Motzkin elimination is then used to obtain the inner bound in terms of $(R_1,R_2,R_3)$. The achievability of the region in \eqref{inner44}--\eqref{inner47} is proved by using the transmission scheme in \eqref{scheme2group4}, and following the same decoding steps as for \eqref{scheme1group4}.
\end{IEEEproof}

\subsection{An Outer Bound}
The outer bound for group 4, stated as Theorem~\ref{theorem:out1}, is formed from the intersection of two outer bounds.
\begin{theorem}\label{theorem:out1} 
If a rate triple $(R_1,R_2,R_3)$ is achievable for a member of group 4, then it must lie in $\mathcal{R}_{\text{out}_1}\cap\mathcal{R}_{\text{out}_2}$ where $\mathcal{R}_{\text{out}_1}$ is the set of all rate triples, each satisfying
\begin{align}
&R_1\leq C\left(\frac{P}{N_1}\right),\label{outer11}\\
&R_2\leq C\left(\frac{(1-\alpha) P}{N_2}\right),\label{outer12}\\
&R_3\leq C\left(\frac{\alpha P}{(1-\alpha) P+N_3}\right),\label{outer13}
\end{align}
for some $0\leq\alpha\leq1$, and $\mathcal{R}_{\text{out}_2}$ is the capacity region of the enhanced channel for the member obtained by decreasing the received noise variance of receiver 3 from $N_3$ to $N_2$.
\end {theorem}
\begin{IEEEproof}
Outer bound 1, $\mathcal{R}_{\text{out}_1}$, follows directly from the outer bound in \eqref{outer561}--\eqref{outer563}, given in the appendix for groups 5 and 6, where $i=2$, $j=3$, and $q=1$ for this group. Outer bound 2, $\mathcal{R}_{\text{out}_2}$, is developed using the idea of enhanced channel. The capacity region of the enhanced channel is an outer bound to the capacity region of the original channel. Since the received noise variance of the two weakest receivers in the defined enhanced channel are equal, we can swap the places of receivers 2 and 3. Then this channel can be considered as a member of group~5 or~3 depending on whether receiver 2 in the original channel knows $M_3$ or not, respectively. Groups~3 and~5 are two of the groups for which we established the capacity region in Section~\ref{section:capacity}.
\end{IEEEproof}

\begin{figure}[t]
	\centering
	\includegraphics[width=0.25\textwidth]{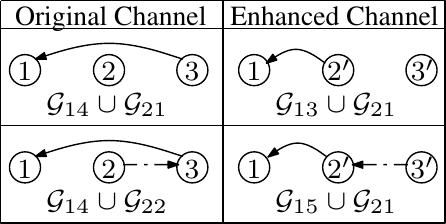}
	\caption{The enhanced channels for $\mathcal{G}_{14}\cup\mathcal{G}_{21}$ and $\mathcal{G}_{14}\cup\mathcal{G}_{22}$, can be considered as $\mathcal{G}_{13}\cup\mathcal{G}_{21}$ and $\mathcal{G}_{15}\cup\mathcal{G}_{21}$, respectively. In the enhanced channel, the places of receivers 2 and 3 can be swapped as they have an equal received noise variance, $N_2$ (the enhanced channel for each member is obtained by decreasing the received noise variance of receiver 3 from $N_3$ to $N_2$).} 
	\label{Fig:enhancedchannel}
\end{figure}

\subsection{Evaluation of the Inner and Outer Bounds} 
In this subsection, we first show that our inner bound for $\mathcal{G}_{14}\cup\mathcal{G}_{2i},\;\,i=2,5$, given in \eqref{inner44}--\eqref{inner47}, is larger than the one achieved by the transmission scheme in \eqref{scheme1group4}, stated as Remark~\ref{remark:group4}. We then show that our outer bound is tighter than the best known outer bound for all the group members. We next characterize the regions where our inner and outer bounds coincide.

\begin{remark}\label{remark:group4}
	Using the transmission scheme in \eqref{scheme1group4}, the region in \eqref{inner41}--\eqref{inner43} is also achievable for $\mathcal{G}_{14}\cup\mathcal{G}_{2i},\;\,i=2,5$. However, the proposed modified scheme in \eqref{scheme2group4} achieves a larger rate region for these two members. To see this, consider any chosen $\alpha$ and $\beta$, conditions~\eqref{inner42} and~\eqref{inner43} are the same as \eqref{inner46} and \eqref{inner47}, but condition~\eqref{inner41} is more restrictive than condition~\eqref{inner44}, and conditions~\eqref{inner41} and \eqref{inner42} are also more restrictive than~\eqref{inner45}.
\end{remark}

\begin{figure}[t]
	\hskip-17pt\includegraphics[width=0.56\textwidth]{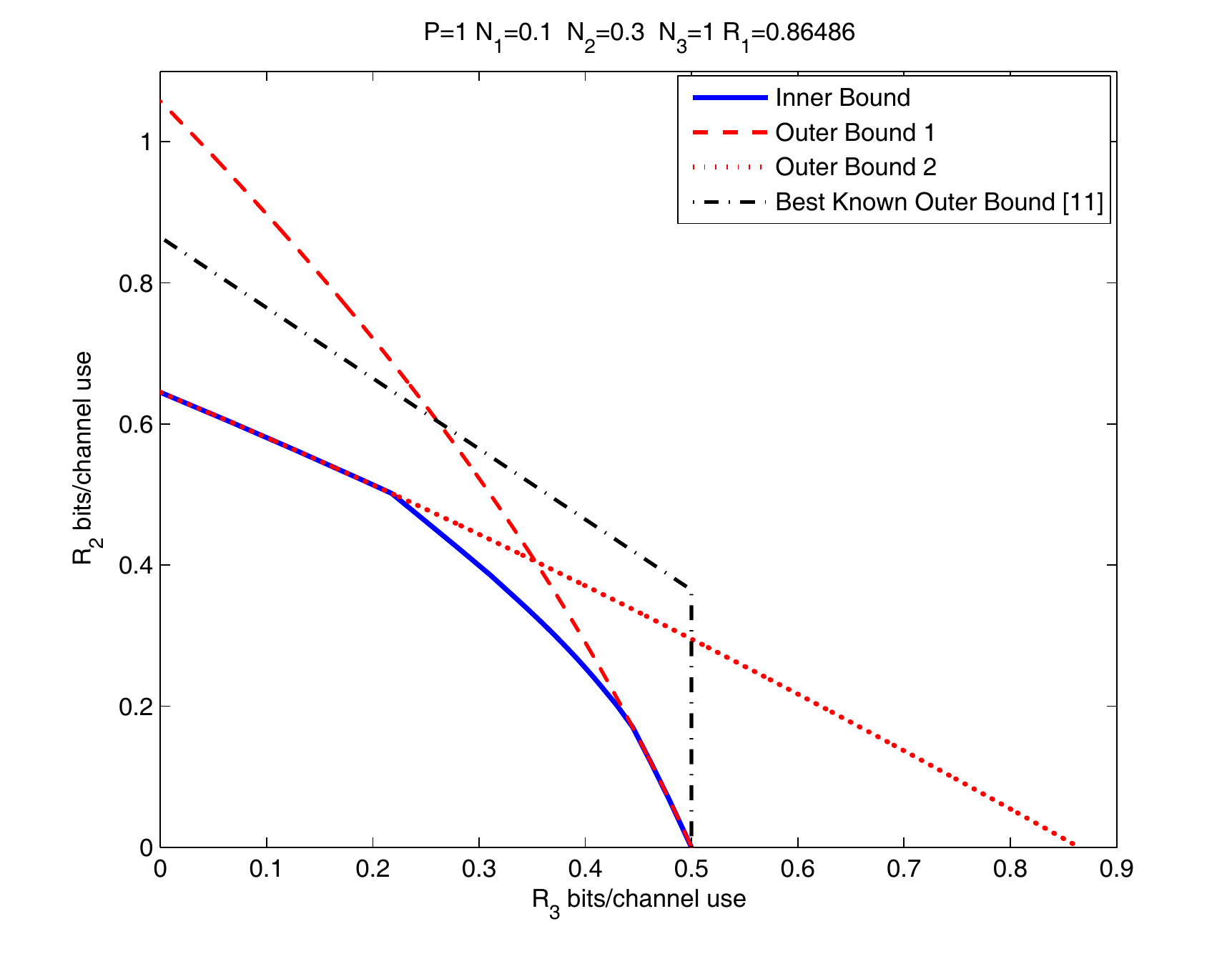}
	\caption{Inner and outer bounds for $\mathcal{G}_{14}\cup\mathcal{G}_{21}$. The proposed outer bound is the intersection of outer bounds 1 and 2.} 
	\label{Group4comparison}
\end{figure}

\begin{figure}[t]
	\centering
	\includegraphics[width=0.23\textwidth]{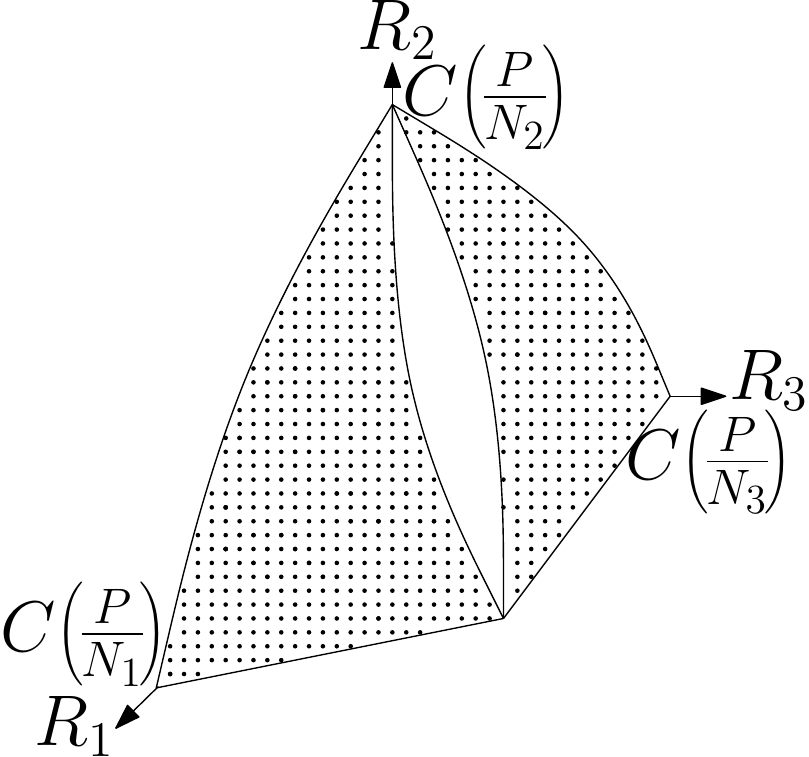}
	\caption{The dotted regions illustrate where the derived inner and outer bounds are tight for $\mathcal{G}_{14}\cup\mathcal{G}_{21}$.} 
	\label{threedimension}
\end{figure}

We now prove that our outer bound is tighter than the best known one, given in~\eqref{bestpriorouterbound}. To this end, we show that, for any condition that must be met in that outer bound, our outer bound includes some more restrictive conditions. We present the proof for $\mathcal{G}_{14}\cup\mathcal{G}_{21}$ in the following; the proof for the other members is similar. Our outer bound is the intersection of the bound given in \eqref{outer11}--\eqref{outer13} and the capacity region of the enhanced channel for $\mathcal{G}_{14}\cup\mathcal{G}_{21}$. The enhanced channel for $\mathcal{G}_{14}\cup\mathcal{G}_{21}$ can be considered as $\mathcal{G}_{13}\cup\mathcal{G}_{21}$ as shown in Fig.~\ref{Fig:enhancedchannel}. According to the enhanced channel, if a rate triple $(R_1,R_2,R_3)$ is achievable for $\mathcal{G}_{14}\cup\mathcal{G}_{21}$, it must satisfy
\begin{align}
R_1+R_3&\leq C\left(\frac{\alpha P}{N_1}\right),\label{outer14}\\
R_2&\leq C\left(\frac{(1-\alpha)P}{\alpha P+N_2}\right),\label{outer15}\\
R_3&\leq C\left(\frac{\alpha P}{N_2}\right),\label{outer16}
\end{align}
for some $0\leq\alpha\leq1$. The best known outer bound, i.e., \eqref{bestpriorouterbound}, states that if a rate triple $(R_1,R_2,R_3)$ is achievable for $\mathcal{G}_{14}\cup\mathcal{G}_{21}$, it must satisfy $R_3\leq C\left(P/N_3\right)$, $R_2+R_3\leq C(P/N_2)$, and $R_1+R_2+R_3\leq C(P/N_1)$. Concerning $R_3\leq C\left(P/N_3\right)$, if condition \eqref{outer13} in $\mathcal{R}_{\text{out}_1}$ is satisfied, this condition is also satisfied.  Conditions \eqref{outer12} and \eqref{outer13} in $\mathcal{R}_{\text{out}_1}$ are more restrictive than $R_2+R_3\leq C(P/N_2)$, and conditions \eqref{outer14} and \eqref{outer15} in $\mathcal{R}_{\text{out}_2}$ are more restrictive than $R_1+R_2+R_3\leq C(P/N_1)$. This completes the proof for $\mathcal{G}_{14}\cup\mathcal{G}_{21}$.

Here we characterize the certain regions where the derived inner and outer bounds coincide. For any fixed $R_1$ where $0\leq R_1\leq C(\frac{P}{N_1})$, the derived bounds are tight when $R_3\leq R_{\text{thr}_3}$ or $R_3\geq R'_{\text{thr}_3}$ where $R_{\text{thr}_3}\hspace{-2pt}\leq\hspace{-2pt}R'_{\text{thr}_3}$; or similarly, when $R_2\leq R_{\text{thr}_2}$ or $R_2\geq R'_{\text{thr}_2}$ where $R_{\text{thr}_2}\leq R'_{\text{thr}_2}$. The thresholds are functions of $R_1$. The same behavior can be observed for any fixed $R_i$ on the $R_j\hspace{-2pt}-\hspace{-2pt}R_q$ plane for any distinct $i,j,q\hspace{-2pt}\in\hspace{-2pt}\{1,2,3\}\hspace{-2pt}$. We present the thresholds on $R_3$ for $\mathcal{G}_{14}\cup\mathcal{G}_{21}$ as an example. For $R_1=0$, $R_{\text{thr}_3}\hspace{-3pt}=\hspace{-3pt}R'_{\text{thr}_3}=0$, and the inner bound when $\beta=0$ coincides with $\mathcal{R}_{\text{out}_1}$. For $0\hspace{-2pt}<\hspace{-2pt}R_1\hspace{-2pt}<\hspace{-2pt}C(\frac{P}{N_1})\hspace{-2pt}-\hspace{-2pt}C(\frac{P}{N_3})$, $R_{\text{thr}_3}\hspace{-3pt}=\hspace{-3pt}C(\frac{\gamma P}{N_3})$ where $\gamma$ satisfies $R_1\hspace{-3pt}=\hspace{-3pt}C(\frac{\gamma P}{N_1})\hspace{-3pt}-\hspace{-3pt}C(\frac{\gamma P}{N_3})$ (in this region, the inner bound when $\beta=1$ coincides with $\mathcal{R}_{\text{out}_2}$), and $R'_{\text{thr}_3}\hspace{-3pt}=\hspace{-3pt}C(\frac{\eta P}{(1-\eta)P+N_3})$ where $\eta$ satisfies $R_1\hspace{-3pt}=\hspace{-3pt}C(\frac{\eta P}{(1-\eta)P+N_1})-C(\frac{\eta P}{(1-\eta)P+N_3})$ (in this region, the inner bound when $\beta=0$ coincides with $\mathcal{R}_{\text{out}_1}$). For $C(\frac{P}{N_1})\hspace{-2pt}-\hspace{-2pt}C(\frac{P}{N_3})\hspace{-2pt}\leq\hspace{-2pt}R_1\hspace{-2pt}<\hspace{-2pt}C(\frac{P}{N_1})$, we have $R_{\text{thr}_3}\hspace{-3pt}=\hspace{-3pt}R'_{\text{thr}_3}=C(\frac{P}{N_1})\hspace{-3pt}-\hspace{-3pt}R_1$, and the inner bound when $\beta=1$ coincides with $\mathcal{R}_{\text{out}_2}$. 

For $\mathcal{G}_{14}\cup\mathcal{G}_{21}$, Fig. \ref{Group4comparison} shows that our outer bound is strictly tighter than best known one given in \eqref{bestpriorouterbound}. This figure also shows that for a fixed $0<R_1<C(\frac{P}{N_1})-C(\frac{P}{N_3})$, the derived bounds coincide when $R_2$ or $R_3$ is below or above certain thresholds. Fig.~\ref{threedimension} also illustrates the behavior of the derived inner and outer bounds in three dimensions.

Comparing with the parallel work by Sima et al. \cite{Capacity3UsersPrivateMessageParallel}, our inner bound is larger than theirs for $G_{14}\cup G_{2i},\;\,i=2,5$, and is the same for the remaining six members. This is because they use the same transmission scheme as in \eqref{scheme1group4} for all the members of the group. Our outer bound is tighter than that by Sima et al. (which coincidentally is also formed from the intersection of multiple outer bounds) for $G_{14}\cup G_{2i},\;\,i=1,3$, and is the same for the remaining six members.

\begin{table*}[t]
	\begin{footnotesize}
		\caption{Group 7: Proposed transmission schemes and inner bounds}
		\vspace{-12pt}
		\begin{center}
			{\renewcommand{\arraystretch}{2}
				\begin{tabular}{|l|l|l|l|}
					\hline
					Member&Graph&Transmission Scheme&Inner Bound ($\mathcal{R}'_{\text{in}}$)\\
					\hline				
					$\mathcal{G}_{17}\cup\mathcal{G}_{21}$&\hspace{-5pt}\raisebox{-0.5ex}{\includegraphics[width=0.1\textwidth]{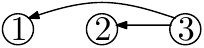}}$\hskip-5pt$&\multirow{2}{*}{$\color{blue}x_1^{(n)}\left([m_{11},m_{31}]\right)+x_2^{(n)}\left([m_2, m_{12},m_{32}]\right)$}&\multirow{1}{*}{\hspace{-4pt}\color{blue}{$R_2+\hspace{-3pt}\underset{i\in\{1,3\}\setminus\mathcal{O}_1}{\sum}\hspace{-5pt}R_i <C\left(\frac{\left(1-\alpha\right)P}{\alpha P+N_2}\right)+C\left(\frac{\alpha P}{N_1}\right),$\hspace{13.5pt}{\tagarray\label{inner7con1}}}}\\
					\cline{1-2}
					$\mathcal{G}_{17}\cup\mathcal{G}_{23}$&\hspace{-5pt}\raisebox{-0.5ex}{\includegraphics[width=0.1\textwidth]{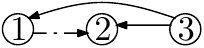}}$\hskip-5pt$&&\multirow{1}{*}{\hspace{47.5pt}\color{blue}{$R_2< C\left(\frac{\left(1-\alpha\right)P}{\alpha P+N_2}\right),$\hspace{54pt}{\tagarray\label{inner7con2}}}}\\
					\cline{1-3}
					$\mathcal{G}_{17}\cup\mathcal{G}_{24}$&\hspace{-5pt}\raisebox{-1ex}{\includegraphics[width=0.1\textwidth]{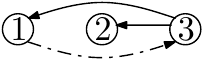}}$\hskip-5pt$&\multirow{2}{*}{$\color{blue}x_1^{(n)}\left([m_{11},m_{31}]\right)+x_2^{(n)}\left([m_2, m_{12}\hspace{-1pt}\oplus\hspace{-1pt}m_{32}]\right)$}&\multirow{1}{*}{\hspace{24.5pt}\color{blue}{$R_2+R_3<C\left(\frac{\left(1-\alpha\right)P}{\alpha P+N_2}\right)+C\left(\frac{\alpha P}{N_3}\right),$\hspace{15pt}{\tagarray\label{inner7con3}}}}\\
					\cline{1-2}
					$\mathcal{G}_{17}\cup\mathcal{G}_{26}$&\hspace{-5pt}\raisebox{-1ex}{\includegraphics[width=0.1\textwidth]{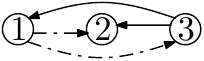}}$\hskip-5pt$&&\multirow{1}{*}{\hspace{47.5pt}\color{blue}{$R_3<C\left(\frac{P}{N_3}\right)$\hspace{74pt}{\tagarray\label{inner7con4}}}}\\
					\hline
					$\mathcal{G}_{17}\cup\mathcal{G}_{22}$&\hspace{-5pt}\raisebox{-0.5ex}{\includegraphics[width=0.1\textwidth]{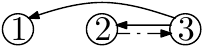}}$\hskip-5pt$&$\color{blue}x_1^{(n)}\left([m_1,m_2\oplus m_{3}]\right)+x_2^{(n)}\left(m_2\oplus m_{3}\right)$&\multirow{1}{*}{\hspace{47.5pt}\color{blue}{$R_1< C\left(\frac{\alpha P}{N_1}\right),$}}\\
					\cline{1-3}
					$\mathcal{G}_{17}\cup\mathcal{G}_{25}$&\hspace{-5pt}\raisebox{-0.5ex}{\includegraphics[width=0.1\textwidth]{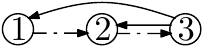}}$\hskip-5pt$&\multirow{3}{*}{$\color{blue}x_1^{(n)}\left([m_1,m_{3}]\right)+x_2^{(n)}\left([m_2,m_{3}]\right)$}&\multirow{1}{*}{\hspace{29.5pt}\color{blue}{$\hspace{-15pt}\underset{i\in\{1,3\}\setminus\mathcal{O}_1}{\sum}\hspace{-5pt}R_i< C\left(\frac{P}{N_1}\right),$}}\\
					\cline{1-2}
					$\mathcal{G}_{17}\cup\mathcal{G}_{27}$&\hspace{-5pt}\raisebox{-1ex}{\includegraphics[width=0.1\textwidth]{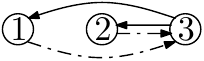}}$\hskip-5pt$&&\multirow{1}{*}{\hspace{47.5pt}\color{blue}{$R_2< C\left(\frac{\left(1-\alpha\right)P}{\alpha P+N_2}\right),$}}\\
					\cline{1-2}
					$\mathcal{G}_{17}\cup\mathcal{G}_{28}$&\hspace{-5pt}\raisebox{-1ex}{\includegraphics[width=0.1\textwidth]{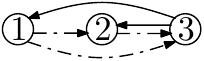}}$\hskip-5pt$&&\multirow{1}{*}{\hspace{47.5pt}\color{blue}{$R_3< C\left(\frac{P}{N_3}\right)\hspace{-5pt}$}}\\
					\hline
			\end{tabular}}
			\label{group7schemes}
		\end{center}
	\end{footnotesize}
\end{table*}

\section{Group 7: Inner and Outer Bounds}\label{Section:Group7}
In this section, we derive inner and outer bounds for group~7, and compare them with the best known ones.

\subsection{Inner Bounds}

\begin{theorem}\label{theorem:in2} 
	A rate triple $(R_1,R_2,R_3)$ for a member of group~7 is achievable if it lies in $\mathcal{R}'_{\text{in}}$, where $\mathcal{R}'_{\text{in}}$ is the set of all rate triples, each satisfying the conditions in the respective row of Table~\ref{group7schemes} for some $0\leq\alpha\leq1$.
\end{theorem}

\begin{IEEEproof}
Our transmission schemes in Table~\ref{group7schemes} are constructed using rate splitting, index coding, multiplexing coding and superposition coding. Each transmission scheme includes two subcodebooks: the first subcodebook consists of i.i.d. codewords generated according to $X_1\sim\mathcal{N}(0,\alpha P)$, and the second subcodebook consists of i.i.d. codewords generated independently according to $X_2\sim\mathcal{N}(0,(1-\alpha)P)$ where $0\leq\alpha\leq1$. For the members using rate splitting, the message $M_i,\;\,i=1,3,$ is divided into two independent messages $\{M_{ij}\}_{j=1}^2$ at rates $\{R_{ij}\}_{j=1}^2$ such that $R_i=R_{i1}+R_{i2}$. For these members, the achievability of $\mathcal{R}'_{\text{in}}$ is verified by employing successive decoding followed by Fourier-Motzkin elimination. At receivers~1 and 3, $x_2^{(n)}$ is first decoded while $x_1^{(n)}$ is treated as noise, and then $x_1^{(n)}$ is decoded. At receiver~2, $x_2^{(n)}$ is just decoded while $x_1^{(n)}$ is treated as noise. For the members not using rate splitting, the achievability of $\mathcal{R}'_{\text{in}}$  is verified by employing simultaneous decoding at receivers~1 and~3, and successive decoding at receiver~2 where $x_2^{(n)}$ is decoded while $x_1^{(n)}$ is treated as noise. For the receivers using simultaneous decoding, the error events can be similarly written as for groups~5 and~6 in Section~\ref{section:capacity}. Note that the receivers utilize their side information during channel decoding.
\end{IEEEproof}

\subsection{An Outer Bound}
The outer bound for group~7, stated as Theorem~\ref{theorem:out2}, is formed from the intersection of two outer bounds. One of them is the best known outer bound given in \eqref{bestpriorouterbound}. 
\begin{theorem}\label{theorem:out2} 
If a rate triple $(R_1,R_2,R_3)$ is achievable for a members of group~7, then it must lie in $\mathcal{R}'_{\text{out}_1}\cap\mathcal{R}'_{\text{out}_2}$ where $\mathcal{R}'_{\text{out}_1}$ is the set of all rate triples, each satisfying
\begin{align}
&R_1\leq C\left(\frac{\alpha P}{N_1}\right),\label{outer21}\\
&R_2\leq C\left(\frac{\left(1-\alpha\right)P}{\alpha P+N_2}\right),\label{outer22}\\
&R_3\leq C\left(\frac{P}{N_3}\right),\label{outer23}
\end{align}
for some $0\leq\hspace{-2pt}\alpha\hspace{-2pt}\leq1$, and $\mathcal{R}'_{\text{out}_2}$ is the outer bound given in \eqref{bestpriorouterbound}.
\end{theorem}
\begin{IEEEproof}
The proof is the same as the converse proof for groups 5 and 6 given in the appendix, where in \eqref{outer561}--\eqref{outer563}, $i=1$, $j=2$, and $q=3$ for this group.
\end{IEEEproof}

\subsection{Evaluation of the Inner and Outer Bounds} 

In this subsection, we show that the derived inner and outer bounds for group 7 coincide for four members and reduce the gap between the best known inner and outer bounds for the remaining four members.

For $\mathcal{G}_{17}\cup\mathcal{G}_{2i},\;\,i=2,5$, the derived outer bound $\mathcal{R}'_{\text{out}_1}\cap\mathcal{R}'_{\text{out}_2}$ coincides with $\mathcal{R}'_{\text{in}}$, which consequently establishes the capacity region. This is while $\mathcal{R}'_{\text{out}_2}$ (the best known outer bound) alone is not tight for these members. 

For $\mathcal{G}_{17}\cup\mathcal{G}_{2i},\;\,i=7,8$, $\mathcal{R}'_{\text{out}_1}$, given in \eqref{outer21}--\eqref{outer23}, coincides with $\mathcal{R}'_{\text{in}}$. This establishes the capacity region for these members and shows that $\mathcal{R}'_{\text{out}_1}$ is strictly tighter than $\mathcal{R}'_{\text{out}_2}$ for these members (this is because $\mathcal{R}'_{\text{out}_1}$ has some curved surfaces while $\mathcal{R}'_{\text{out}_2}$ is a polyhedron).

For the remaining four members, we now show that our inner and outer bounds are both strictly tighter than the best known ones. For these members, we first improve the best known inner bound given in~\eqref{deterministicregion} by using the joint decoding approach. We then show that our inner bound in Theorem~\ref{theorem:in2} is even larger than the inner bound achieved by this approach. Using the joint decoding approach, the inner bound for the remaining four members with unknown capacity region is the set of all rate triples $(R_1,R_2,R_3)$, each satisfying
\begin{align}
R_2+\sum_{i\in\{1,3\}\setminus\mathcal{O}_1}{R_i}&<B_1+B_2+B'_3,\label{bestpriorinner1}\\
R_2+R_3&<B_2+B'_3,\label{bestpriorinner2}\\
R_3&<\min\{C\left(\frac{\alpha_3P}{N_3}\right),B'_3\},\label{bestpriorinner3}
\end{align}
for some $\alpha_\ell\geq0$ such that $\sum_{\ell=1}^{3}{\alpha_\ell}=1$ ($B_1$, $B_2$ and $B'_3$ are defined the same as in Section~\ref{section:looseness}).
This inner bound for $\mathcal{G}_{17}\cup\mathcal{G}_{2i},\;\,i=1,3,$ is achieved using the scheme
\begin{equation*}
x_1^{(n)}(m_{11})+x_2^{(n)}([m_{12}, m_{21}])+x_3^{(n)}([m_{13},m_{22},m_3]),
\end{equation*}
and for $\mathcal{G}_{17}\cup\mathcal{G}_{2i},\;\,i=4,6,$ using the scheme
\begin{equation*}
x_1^{(n)}(m_{11})+x_2^{(n)}([m_{12}, m_{21}])+x_3^{(n)}([m_{13}\hspace{-2pt}\oplus\hspace{-2pt}m_3,m_{22}]),
\end{equation*}
where the three subcodebooks are constructed independently using i.i.d. codewords generated according to $X_\ell\hspace{-2pt}\sim\hspace{-2pt}\mathcal{N}(0,\alpha_\ell P)$ for some $\alpha_\ell\geq0$ such that $\sum_{\ell=1}^{3}\alpha_\ell=1$. 

\begin{figure}[t]
	\hskip-17pt\includegraphics[width=0.56\textwidth]{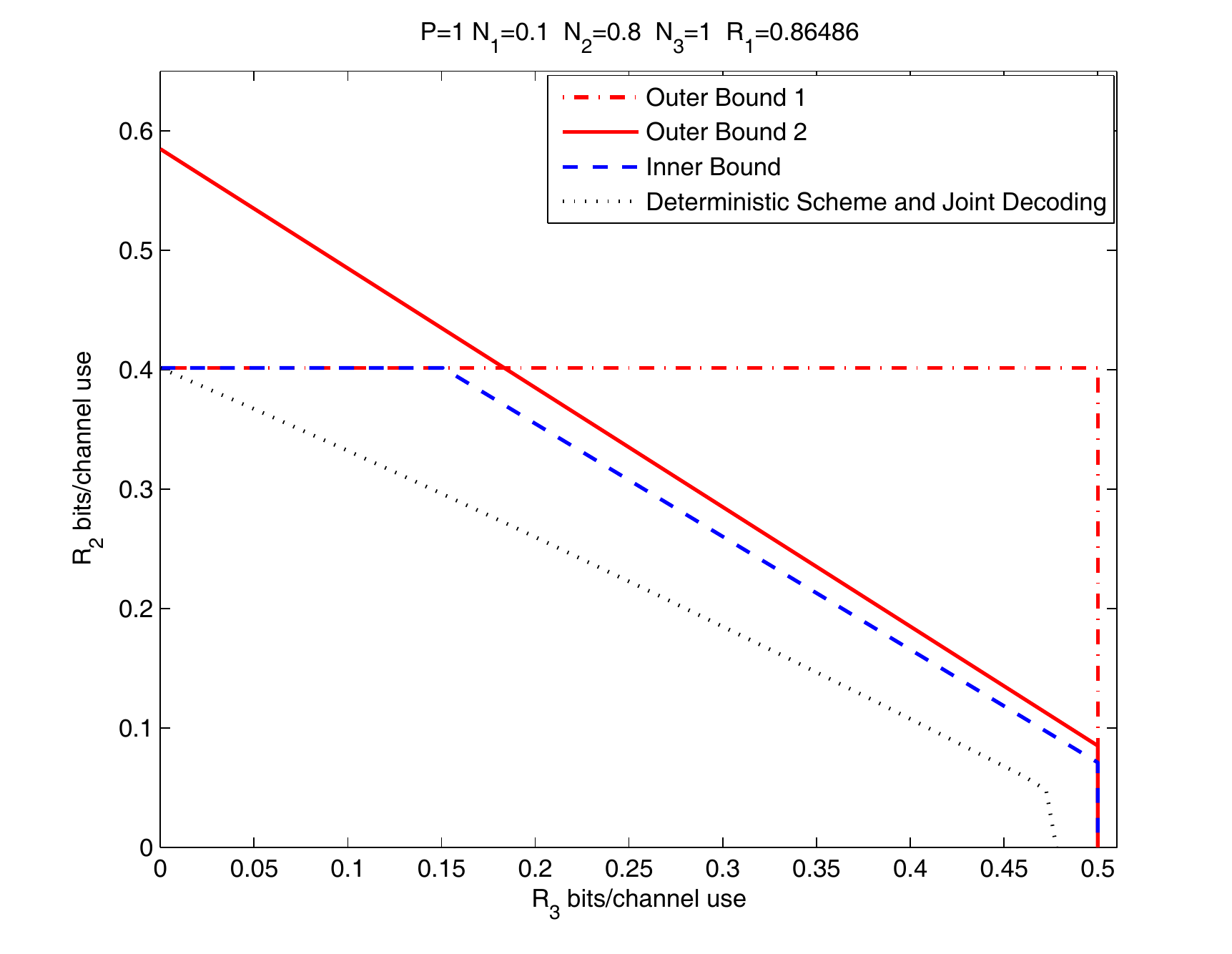}
	\caption{Inner bound and outer bound comparison for $\mathcal{G}_{17}\cup\mathcal{G}_{24}$.} 
	\label{Group7comparison}
\end{figure}

We now show that for any chosen set of $\{\alpha_\ell\}_{\ell=1}^3$, the region in \eqref{bestpriorinner1}--\eqref{bestpriorinner3} is smaller than $\mathcal{R}'_{\text{in}}$ for $\alpha=\alpha_1$. Noting that $B_2+B'_3=C\left(\frac{(1-\alpha_1)P}{\alpha_1 P+N_2}\right)$, then {condition~\eqref{inner7con1}} in $\mathcal{R}'_{\text{in}}$ is the same as \eqref{bestpriorinner1}, {conditions~\eqref{inner7con2} and~\eqref{inner7con3}} are more relaxed than \eqref{bestpriorinner2}, and {condition~\eqref{inner7con4}} is more relaxed than \eqref{bestpriorinner3}. This proves that our inner bound in Theorem~\ref{theorem:in2} is larger than the inner bound in \eqref{bestpriorinner1}--\eqref{bestpriorinner3}.

Concerning the outer bound, since our outer bound is the intersection of the best known outer bound and a new outer bound, $\mathcal{R}'_{\text{out}_1}$, hence, it is tighter than the best known outer bound. As an example, for $\mathcal{G}_{17}\cup\mathcal{G}_{24}$, Fig. \ref{Group7comparison} depicts that our inner and outer bounds are strictly tighter than the best known inner and outer bounds.

Comparing with the parallel work by Sima et al. \cite{Capacity3UsersPrivateMessageParallel}, their inner and outer bounds are tight only for $G_{17}\cup G_{2i},\;\,i=7,8$, while our inner and outer bounds are tight for $G_{17}\cup G_{2i},\;\,i=2,5,$ as well. Our outer bound is the same as theirs for all the members of the group, which indicates that our inner bound is larger than theirs for $G_{17}\cup G_{2i},\;\,i=2,5$. For the remaining four members with unknown capacity region, we were not able to prove that our inner bound is larger. However, by numerical simulation with different channel parameters, we found that our inner bound was larger than their inner bound for all the cases we considered.

\begin{table*}[t]
		\begin{footnotesize}
			\caption{Members using Index Coding and their transmission schemes} 
			\vspace{-13pt}
			\begin{center}
				{\renewcommand{\arraystretch}{2}
					\begin{tabular}{|l|l|l|}
						\hline
						Member & Graph & Transmission Scheme\\
						\hline
						$\mathcal{G}_{12}\cup\mathcal{G}_{22}$&\raisebox{-0.5ex}{\includegraphics[width=0.1\textwidth]{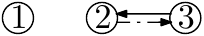}}&$\color{blue}x_1^{(n)}\left(m_1\right)\hspace{-2pt}+\hspace{-2pt}x_2^{(n)}\left(m_2\oplus m_3\right)$\\
						\hline
						$\mathcal{G}_{15}\cup\mathcal{G}_{22}$&\raisebox{-0.5ex}{\includegraphics[width=0.1\textwidth]{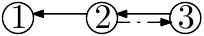}}&$\color{blue}x_1^{(n)}\left([m_1,m_2\oplus m_3]\right)\hspace{-2pt}+\hspace{-2pt}x_2^{(n)}\left(m_2\oplus m_3\right)$\\
						\hline
						$\mathcal{G}_{18}\cup\mathcal{G}_{22}$&\raisebox{-0.5ex}{\includegraphics[width=0.1\textwidth]{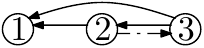}}&$\color{blue}x^{(n)}\left([m_1,m_2\oplus m_3]\right)$\\
						\hline
						$\mathcal{G}_{14}\cup\mathcal{G}_{22}$&\raisebox{-0.5ex}{\includegraphics[width=0.1\textwidth]{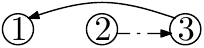}}&\multirow{2}{*}{$\color{blue}x_1^{(n)}\left([m'_{1}\oplus m'_{31},m'_{32}],x_2^{(n)}\right)+x_2^{(n)}\left([m_2,m_{31}],x_3^{(n)}\right)+x_3^{(n)}\left([m''_{1}\oplus m''_{31},m''_{32}]\right)$}\\
						\cline{1-2}
						$\mathcal{G}_{14}\cup\mathcal{G}_{25}$&\raisebox{-0.5ex}{\includegraphics[width=0.1\textwidth]{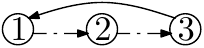}}&\\
						\hline
						$\mathcal{G}_{17}\cup\mathcal{G}_{22}$&\raisebox{-0.5ex}{\includegraphics[width=0.1\textwidth]{FiguresBCwithSI/G17G22.pdf}}&$\color{blue}x_1^{(n)}\left([m_1,m_2\oplus m_3]\right)\hspace{-2pt}+\hspace{-2pt}x_2^{(n)}\left(m_2\oplus m_3\right)\hspace{-4pt}$\\
						\hline
						$\mathcal{G}_{17}\cup\mathcal{G}_{24}$&\raisebox{-1ex}{\includegraphics[width=0.1\textwidth]{FiguresBCwithSI/G17G24.pdf}}&\multirow{2}{*}{$\color{blue}x_1^{(n)}\left([m_{11},m_{31}]\right)+x_2^{(n)}\left([m_2, m_{12}\hspace{-3pt}\oplus\hspace{-2pt}m_{32}]\right)$}\\
						\cline{1-2}
						$\mathcal{G}_{17}\cup\mathcal{G}_{26}$&\raisebox{-1ex}{\includegraphics[width=0.1\textwidth]{FiguresBCwithSI/G17G26.pdf}}&\\
						\hline
				\end{tabular}}
			\label{memberswithindexcoding}
		\end{center}
	\end{footnotesize}
\end{table*}

\section{Remarks on the Transmission Schemes}
In this section, we provide some remarks on index coding, dirty paper coding, and simultaneous decoding, which highlight the roles that these coding techniques play in our transmission schemes. 
We then extend our methodology to AWGN broadcast channels with more than three receivers and conjecture the groups for which we can establish the capacity region for all the group members.

\subsection{Index Coding}\label{subsection:discussindex}
Index coding has been used in the transmission schemes of eight side information configurations, shown in Table~\ref{memberswithindexcoding}. If we replace index coding with multiplexing coding for these members, we cannot achieve the same region. This observation is consistent with our previous study on the non-interchangeability of index and multiplexing coding for some side information configurations in AWGN broadcast channels with three or more receivers \cite{IndexCodingvsMultiplexingCoding}. In all of these transmission schemes, there exists a receiver who decodes the XOR of two messages, both not known a priori. From the standpoint of this receiver, replacing index coding with multiplexing coding increases the amount of uncertainty to be resolved in the resultant message from $n\max\{R_i,R_j\}$ to $n(R_i+R_j)$ bits.

\subsection{Dirty Paper Coding}
The transmission schemes for group~4 (transmission schemes \eqref{scheme1group4} and \eqref{scheme2group4}) employ dirty paper coding. In this group, $M_1$ (i.e., the message intended for the strongest receiver) and $M_3$ (i.e., the message intended for the weakest receiver) are multiplexed, and receiver~2 (the in-between receiver) does not know $M_1$. At each receiver, dirty paper coding allows us to avoid decoding a message intended for a stronger receiver (which otherwise imposes additional rate constraints), and to cancel some parts of the signal carrying a message intended for a weaker receiver (which increases the signal-to-noise ratio). For example, at receiver~2, we avoid decoding $M_1$, and at the same time we cancel a part of the signal which carries $M_3$, i.e., $x^{(n)}_3$.

Dirty paper coding can also be used to provide a unified scheme for group~2, avoiding the exception $\mathcal{G}_{12}\cup\mathcal{G}_{22}$ in that group. The following transmission scheme can achieve the capacity region for all the group members,
\begin{align}\label{group2DPC}
x_1^{(n)}\left(m_1,x_2^{(n)}\right)+x_2^{(n)}\left([m_2,m_3]\right).
\end{align}
In this scheme, we consider $x_2^{(n)}$ as interference for receiver~1 known non-causally at the transmitter. The first subcodebook is then constructed using dirty paper coding with the auxiliary random variable $U_1=X_1+\lambda_1 X_2$ where $X_1\sim\mathcal{N}\left(0,\alpha P\right)$ is independent of $X_2\sim\mathcal{N}\left(0,(1-\alpha)P\right)$, and $\lambda_1=\frac{\alpha P}{\alpha P+N_1}$. To achieve the capacity region, receivers~2 and 3 treat $x_1^{(n)}$ as noise, and decode $x_2^{(n)}$ using their side information. Then receiver~2 can reliably decode $M_2$ if $\sum_{i\in\{2,3\}\setminus\mathcal{O}_2} {R_i}< C(\frac{\alpha_2P}{\alpha_1P+N_2})$, and receiver~3 can reliably decode $M_3$ if $R_3< C(\frac{\alpha_2P}{\alpha_1P+N_3})$. Receiver~1 decodes $m_1$ without being affected by $x_2^{(n)}$ due to dirty paper coding. Then this receiver, irrespective of its side information, can reliably decode $M_1$ if $R_1<C(\frac{\alpha_1 P}{N_1})$. Using \eqref{group2DPC}, since we need not decode $x_2^{(n)}$ at receiver~1, there is no longer any need for the index coding of $m_2$ and $m_3$ wherever possible.

\subsection{Simultaneous Decoding}
Simultaneous decoding is utilized for groups~5 and~6, and four members of group~7. For each of these side information configurations, there is at least one message that is an operand in both subcodebooks of the transmission scheme. As an example, for $\mathcal{G}_{15}\cup\mathcal{G}_{21}$, the two subcodebooks of the transmission scheme are both functions of $m_2$. For group 6, simultaneous decoding can be replaced with rate splitting and successive decoding. For this group, we can also achieve the capacity region by using the alternative transmission scheme
\begin{align*}
	x_1^{(n)}\left([m_{11},m_2]\right)+x_2^{(n)}\left([m_{12},m_3]\right),
\end{align*}
where rate splitting is utilized, and using successive decoding alone (where, at receivers~1 and~2, $x_2^{(n)}$ is first decoded while $x_1^{(n)}$ is treated as noise, and then $x_1^{(n)}$ is decoded. At receiver~3, $x_2^{(n)}$ is only decoded while $x_1^{(n)}$ is treated as noise).

For groups~5 and~7, as opposed to group~6, there exist some cases for which simultaneous decoding cannot similarly be replaced with rate splitting and successive decoding. As an example, consider group~5's leader, $\mathcal{G}_{15}\cup\mathcal{G}_{21}$. By utilizing the same approach as for group~6, suppose that we use the transmission scheme
\begin{align*}
	x_1^{(n)}\left([m_{1},m_{21}]\right)+x_2^{(n)}\left([m_{22},m_3]\right),
\end{align*}
and employ successive decoding (where, at receivers~1 and~2, $x_2^{(n)}$ is first decoded while $x_1^{(n)}$ is treated as noise, and then $x_1^{(n)}$ is decoded. At receiver~3, $x_2^{(n)}$ is only decoded while $x_1^{(n)}$ is treated as noise). This leads to a strictly smaller achievable rate region when $N_1<N_2$. We can verify that other decoding orders (of successive decoding) also lead to suboptimal results.

Our use of simultaneous decoding has helped us establish the capacity region for six more side information configurations (in groups~5 and~7) compared to the parallel work by Sima et al.~\cite{Capacity3UsersPrivateMessageParallel} (in which rate splitting and successive decoding are utilized, but not simultaneous decoding).

\subsection{AWGN Broadcast Channels with More Than Three Receivers}
In this subsection, we consider private-message broadcasting over AWGN broadcast channels with $Q>3$ receivers where without loss of generality $N_1\leq N_2\leq N_3\leq\cdots\leq N_Q$. We perform the same grouping method as for the three-receiver case. In our proposed classification, all the side information graphs with the same subgraph obtained by removing all the arcs from a stronger receiver to a weaker receiver, are classified into one group; the resultant common subgraph for each group is considered as the group leader. We then, using the insights gained from the three-receiver case, conjecture a condition that we can use to characterize some groups for which we can establish the capacity region for all the members. 
\begin{conjecture}\label{conjecture1}
If for every receiver $i$ in a group leader, each side information message $M_j$ (i.e., $M_j\in\mathbf{K}_i$) is also known to all the receivers that are (i) stronger than receiver $i$, and (ii) weaker than receiver $j$ (i.e., all receivers $q$ where $j<q<i$), we can establish the capacity region for all the group members. The capacity region is achieved using \textit{Codebook Construction~A} and \textit{Decoding Scheme~A} which are defined as follows.
\end{conjecture}

\textit{Codebook Construction A}: The transmission scheme for a group leader is constructed using the following method. A subcodebook is constructed associated with a receiver~$i$ if there is at least one message in $M_i\cup\mathbf{K}_i$ that is not known a priori to any weaker receiver. This condition is true even when (i) $\mathbf{K}_i=\emptyset$ and $M_i$ is not known a priori to any weaker receiver, or (ii) receiver~$i$ is the weakest receiver, i.e., $i=Q$. This subcodebook is formed by multiplexing $M_i$ with $\mathbf{K}_i$. The transmission scheme is finally formed from the linear superposition of the subcodebooks. For some members of the group which are not the group leader, index coding may also be required following the form outlined in Section~\ref{subsection:discussindex}.

\textit{Decoding scheme A}: At each receiver $i$, simultaneous decoding is performed over all the subcodebooks that contain only $M_i$, messages intended for weaker receivers (i.e., $M_j$, $j>i$), or messages known to the receiver.

\begin{figure}[b]
	\centering
	\includegraphics[width=0.165\textwidth]{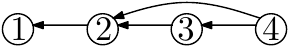}
	\caption{A group leader for the four-receiver AWGN broadcast channel with receiver message side information.}
	\label{FourReceiverConfiguration}
\end{figure}

As an example, we here establish the capacity region for a group leader of the four-receiver AWGN broadcast channel with receiver message side information shown in Fig.~\ref{FourReceiverConfiguration}.

\begin{theorem}\label{capacityfoureceiver}
	The capacity region of the four-receiver AWGN broadcast channel with the side information configuration shown in Fig. \ref{FourReceiverConfiguration} is the closure of the set all rate tuples $(R_1,R_2,R_3,R_4)$, each satisfying
	\begin{align}
	R_1&<C\left(\frac{\alpha P}{N_1}\right),\label{4receiver1}\\
	R_1+R_2+R_3+R_4&<C\left(\frac{P}{N_1}\right),\label{4receiver2}\\
	R_2+R_3+R_4&<C\left(\frac{P}{N_2}\right),\label{4receiver3}\\
	R_3+R_4&<C\left(\frac{(1-\alpha)P}{\alpha P+N_3}\right),\label{4receiver4}\\
	R_4&<C\left(\frac{(1-\alpha)P}{\alpha P+N_4}\right)\label{4receiver5},
	\end{align}
	for some $0\leq\alpha\leq1$.
\end{theorem}
\begin{IEEEproof}
(\textit{Codebook Construction}) Using \textit{Codebook Construction A}, the capacity-achieving transmission scheme takes the form
\begin{align*}
x_1^{(n)}\left([m_1,m_2]\right)+x_2^{(n)}\left([m_2,m_3,m_4]\right),
\end{align*}
where the first subcodebook consists of i.i.d. codewords generated according to $X_1\sim\mathcal{N}\left(0,\alpha P\right)$ for each $[m_1,m_2]$, the second subcodebook consists of i.i.d. codewords generated according to $X_2\sim\mathcal{N}\left(0,(1-\alpha) P\right)$ for each $[m_2,m_3,m_4]$, and $X_1$ is independent of $X_2$.

(\textit{Achievability Proof}) At receivers~1 and~2, simultaneous decoding is performed over $x_1^{(n)}+x_2^{(n)}$, and at receivers~3 and~4, $x_2^{(n)}$ is decoded while $x_1^{(n)}$ is treated as noise. Receiver~1 can reliably decode $M_1$ if conditions~\eqref{4receiver1} and~\eqref{4receiver2} hold. Receiver~2 can reliably decode $M_2$ if condition~\eqref{4receiver3} holds. Receiver~3 can reliably decode $M_3$ if condition~\eqref{4receiver4} holds, and receiver~4 can reliably decode $M_4$ if condition~\eqref{4receiver5} holds (the error events are similarly written as for groups~5 and~6 in Section~\ref{section:capacity}).
	
(\textit{Converse Proof}) We just define $\alpha$ for this side information configuration, and the remainder of the converse proof uses the same techniques introduced for group 2 in the appendix. For this side information configuration, $R_3+R_4$ is upper bounded~as
\begin{align*}
&\hskip-5pt n(R_3+R_4)\nonumber\\
&=H(M_3,M_4 \mid Y_3^{(n)}, M_2)+I(M_3,M_4;Y_3^{(n)}\mid M_2)\nonumber\\
&\overset{(a)}{\leq}2 n\epsilon_n+h(Y_3^{(n)}\mid M_2)-h(Y_3^{(n)}\mid M_2,M_3,M_4)\nonumber\\
&\leq2 n\epsilon_n+\frac{n}{2}\log2\pi e(P+N_3)-h(Y_3^{(n)}\mid M_2,M_3,M_4) \nonumber\\
&\overset{(b)}{=}2 n\epsilon_n+\frac{n}{2}\log2\pi e(P+N_3)-\frac{n}{2}\log2\pi e(\alpha P+N_3),
\end{align*}
for some $0\leq\alpha\leq1$ where $(a)$ and $(b)$ follow from similar reasons as $(a)$ and $(b)$ in \eqref{proof21}.
\end{IEEEproof}

\section{Conclusion}
We considered the problem of private-message broadcasting over the three-receiver AWGN broadcast channel with receiver message side information. We first classified all 64 possible side information configurations into eight groups, each consisting of eight members. We then derived inner and outer bounds for different groups, and established the capacity region for 52 out of 64 possible side information configurations. For six groups, i.e., all the groups except groups~4 and~7, we established the capacity region for all the group members, and proved the looseness of both the best known inner and outer bounds. For group~4, we improved the best known inner bound and/or outer bound for all the group members. For this group, our bounds coincide at certain regions, which can be characterized by two thresholds. For group~7, we established the capacity region for four members. For the remaining four members, we proved that our inner and outer bounds are both tighter than the best known inner and outer bounds.

\section*{Appendix}
In this section, we prove the converse part of Theorem~\ref{maintheorem}. The proof is based on those for AWGN broadcast channels without side information~\cite{AWGNBCConverse, NITBook}. In the converse, we use Fano's inequality and the entropy power inequality (EPI). Based on Fano's inequality, we have
\begin{equation}
H(M_i \mid Y_i^{(n)},\mathbf{K}_i)\leq n\epsilon_{n,i},\quad i = 1,2,3\label{fano},
\end{equation}
where $\epsilon_{n,i} \rightarrow 0$ as $n\rightarrow \infty$. For the sake of simplicity, we use $\epsilon_n$ instead of $\epsilon_{n,i}$ for the remainder. We also use the fact that the capacity region of a stochastically degraded broadcast channel without feedback is the same as its equivalent physically degraded broadcast channel~\cite[p. 444]{NITBook} for which we have the Markov chain $(M_1,M_2,M_3)\rightarrow X\rightarrow Y_{1}\rightarrow Y_{2} \rightarrow  Y_{3}$, i.e., $Y_1=X+Z_1$ and $Y_i=Y_{i-1}+\tilde{Z}_i, \;\;i=2,3$, where $\tilde{Z}_i \sim \mathcal{N}\left(0, N_i-N_{i-1}\right)$.  

\begin{IEEEproof} we present the converse proof for different groups.

\textit{Group 1}: The converse proof for all the members of this group is the same as the group leader, i.e., the three-receiver AWGN broadcast channel without receiver message side information for which the converse proof exists \cite{AWGNBCConverse}.
%****************************************************

\textit{Group 2}: For the members where receiver~2 does not know $M_3$ as side information, $R_2+R_3$ is upper bounded as
\begin{align}\label{proof21}
&\hskip-4pt n(R_2+R_3)\nonumber\\
&=H(M_2,M_3 \mid Y_2^{(n)})+I(M_2,M_3;Y_2^{(n)})\nonumber\\
&\overset{(a)}{\leq}2 n\epsilon_n+h(Y_2^{(n)})-h(Y_2^{(n)}\mid M_2,M_3)\nonumber\\
&\leq2 n\epsilon_n+\frac{n}{2}\log2\pi e(P+N_2)-h(Y_2^{(n)}\mid M_2,M_3) \nonumber\\
&\overset{(b)}{=}2 n\epsilon_n\hskip-2pt+\hskip-2pt\frac{n}{2}\log2\pi e(P+N_2)\hskip-2pt-\hskip-2pt\frac{n}{2}\log2\pi e(\alpha P+N_2),
\end{align}
for some $0\leq\alpha\leq1$ where $(a)$ follows from adding the following inequalities resulted from the physical degradedness of the channel and \eqref{fano}.
\begin{align*}
	H(M_3\mid Y_2^{(n)},M_2)&\leq H(M_3\mid Y_3^{(n)},M_2)  \leq n\epsilon_n,\\
 H(M_2\mid Y_2^{(n)}) & \leq n \epsilon_n.
\end{align*}
In \eqref{proof21}, $(b)$ follows from
\begin{multline*}
\frac{n}{2}\log2\pi e N_2=h(Z_2^{(n)})=h(Y_2^{(n)}\hspace{-4pt}\mid\hspace{-4pt}X^{(n)})\\ \leq h(Y_2^{(n)}\mid M_2,M_3)\leq \frac{n}{2}\log2\pi e(P+N_2),
\end{multline*}
then since 
\begin{align*}
\frac{n}{2}\log2\pi e N_2\leq h(Y_2^{(n)}\mid M_2,M_3)\leq \frac{n}{2}\log2\pi e(P+N_2), 
\end{align*}
there must exist an $0\leq \alpha \leq1$ such that 
\begin{align}
& h(Y_2^{(n)}\mid M_2,M_3)=\frac{n}{2}\log2\pi e(\alpha P+N_2)\label{alpha2}.
\end{align}
For these members, we also have
\begin{align}\label{proof22}
&\hskip-4pt nR_3\nonumber\\
&=H(M_3 \mid Y_3^{(n)},M_2)+I(M_3;Y_3^{(n)}\mid M_2)\nonumber\\
&\overset{(a)}{\leq} n\epsilon_n+h(Y_3^{(n)}\mid M_2)-h(Y_3^{(n)}\mid M_2,M_3)\nonumber\\
&\leq n\epsilon_n+\frac{n}{2}\log2\pi e(P+N_3)-h(Y_3^{(n)}\mid M_2,M_3) \nonumber\\
&\overset{(b)}{\leq}n\epsilon_n\hskip-2pt+\hskip-2pt\frac{n}{2}\log2\pi e(P+N_3)\hskip-2pt-\hskip-2pt\frac{n}{2}\log2\pi e(\alpha P+N_3),
\end{align}
and
\begin{align}\label{proof23}
&\hskip-4pt nR_1\nonumber\\
&=H(M_1 \mid Y_1^{(n)},M_2,M_3)+I(M_1;Y_1^{(n)} \mid M_2,M_3) \nonumber\\
&\overset{(a)}{\leq} n\epsilon_n+h(Y_1^{(n)}\mid M_2,M_3)-h(Y_1^{(n)}\mid M_1,M_2,M_3) \nonumber\\
&\overset{(c)}{\leq} n\epsilon_n+\frac{n}{2}\log2\pi e(\alpha P+N_1)-h(Y_1^{(n)}\mid M_1,M_2,M_3) \nonumber\\
&\overset{(d)}{=}n\epsilon_n+\frac{n}{2}\log2\pi e(\alpha P+N_1)-\frac{n}{2}\log2\pi e N_1,
\end{align}
where $(a)$ follows from \eqref{fano}, $(b)$ from using the conditional EPI {\cite[p. 22]{NITBook}} for $Y_3^{(n)}=Y_{2}^{(n)}+\tilde{Z}_3^{(n)}$, and substituting $h(\tilde{Z}_3^{(n)}\mid M_2,M_3)=\frac{n}{2}\log2\pi e(N_3-N_2)$ and \eqref{alpha2}, $(c)$ from using the conditional EPI for $Y_2^{(n)}=Y_{1}^{(n)}+\tilde{Z}_2^{(n)}$, and substituting $h(\tilde{Z}_2^{(n)}\mid M_2,M_3)=\frac{n}{2}\log2\pi e(N_2-N_1)$ and \eqref{alpha2}, and $(d)$ from
\begin{multline}\label{noiseentropy}
h(Y_i^{(n)}\mid M_1,M_2,M_3)=h(Y_i^{(n)}\mid X^{(n)})=\\h(Z_i^{(n)})=\frac{n}{2}\log 2\pi eN_i, \hskip5pt i=1,2,3.
\end{multline}

From \eqref{proof21}, \eqref{proof22}, \eqref{proof23} and since $\epsilon_n$ goes to zero as $n \rightarrow \infty$, the converse proof for the members where receiver~2 does not know $M_3$ a priori is complete. For the members where receiver~2 knows $M_3$ a priori, we just need to modify \eqref{proof21} as follows, and reuse \eqref{proof22} and \eqref{proof23}.
\begin{align*} 
&\hskip-5pt nR_2\nonumber\\
&=H(M_2 \mid Y_2^{(n)},M_3)+I(M_2;Y_2^{(n)} \mid M_3)\nonumber\\
&\overset{(a)}{\leq} n\epsilon_n+h(Y_2^{(n)}\mid M_3)-h(Y_2^{(n)}\mid M_2,M_3) \nonumber\\
&\leq n\epsilon_n+\frac{n}{2}\log2\pi e(P+N_2)-h(Y_2^{(n)}\mid M_2,M_3) \nonumber\\
&\overset{(b)}{=}n\epsilon_n+\frac{n}{2}\log2\pi e(P+N_2)-\frac{n}{2}\log2\pi e(\alpha P+N_2),
\end{align*}
where $(a)$ follows from \eqref{fano}, and $(b)$ from the same reason as $(b)$ in \eqref{proof21}.
%******************************************************************************

\textit{Group 3}: For this group, we just define $\alpha$, and the remainder of the converse proof uses the same techniques employed for group~2. In this group, $R_3$ is upper bounded as
\begin{align}\label{proof31}
&\hskip-4pt nR_3\nonumber\\
&=H(M_3 \mid Y_3^{(n)})+I(M_3;Y_3^{(n)})\nonumber\\
&\overset{(a)}{\leq} n\epsilon_n+h(Y_3^{(n)})-h(Y_3^{(n)}\mid M_3) \nonumber\\
&\leq n\epsilon_n+\frac{n}{2}\log2\pi e(P+N_3)-h(Y_3^{(n)}\mid M_3) \nonumber\\
&\overset{(b)}{=}n\epsilon_n\hskip-2pt+\hskip-2pt\frac{n}{2}\log2\pi e(P+N_3)\hskip-2pt-\hskip-2pt\frac{n}{2}\log2\pi e(\alpha P+N_3),
\end{align}
for some $0\leq\alpha\leq1$ where $(a)$ follows from \eqref{fano}, and $(b)$ from similar reason as $(b)$ in \eqref{proof21}, i.e., since
\begin{align*}
\frac{n}{2}\log2\pi e N_3\leq h(Y_3^{(n)}\mid M_3)\leq \frac{n}{2}\log2\pi e(P+N_3),
\end{align*}
there must exist an $0\leq \alpha \leq1$ such that 
\begin{align}
h(Y_3^{(n)}\mid M_3)=\frac{n}{2}\log2\pi e(\alpha P+N_3).\label{alpha3}
\end{align}
%******************************************************************************

\textit{Groups 5 and 6:} The outer bound for these two groups is formed from the intersection of the following outer bound and the best known outer bound given in \eqref{bestpriorouterbound}.
If a rate triple $(R_1,R_2,R_3)$ is achievable for a member of groups~5 and~6, then it must satisfy
\begin{align}
	&R_i\leq C\left(\frac{\alpha P}{N_i}\right),\label{outer561}\\
	&R_j\leq C\left(\frac{\left(1-\alpha\right)P}{\alpha P+N_j}\right),\label{outer562}\\
	&R_k\leq C\left(\frac{P}{N_q}\right),\label{outer563}
\end{align}
for some $0\leq\hspace{-2pt}\alpha\hspace{-2pt}\leq1$, where $i=1$, $j=3$, and $q=2$ for group 5, and $i=2$, $j=3$, and $q=1$ for group 6.
Conditions \eqref{outer561} and \eqref{outer562} follow from the capacity region of the two-receiver AWGN broadcast channel (from the transmitter to receivers~$i$ and~$j$) where only the stronger receiver (receiver~$i$) may know the requested message of the weaker receiver (receiver~$j$) as side information. The side information of receivers~$i$ and~$j$ about each other's requested messages has this property. Condition \eqref{outer563} is due to the point-to-point channel capacity between the transmitter and receiver~$q$.
%******************************************************************************

\textit{Group 8}: Best known outer bound given in \eqref{bestpriorouterbound} is tight.
\end{IEEEproof}
\bibliographystyle{IEEEtran}
%\bibliography{IEEEabrv,BCwithSI}

\begin{thebibliography}{10}
	\providecommand{\url}[1]{#1}
	\csname url@samestyle\endcsname
	\providecommand{\newblock}{\relax}
	\providecommand{\bibinfo}[2]{#2}
	\providecommand{\BIBentrySTDinterwordspacing}{\spaceskip=0pt\relax}
	\providecommand{\BIBentryALTinterwordstretchfactor}{4}
	\providecommand{\BIBentryALTinterwordspacing}{\spaceskip=\fontdimen2\font plus
		\BIBentryALTinterwordstretchfactor\fontdimen3\font minus
		\fontdimen4\font\relax}
	\providecommand{\BIBforeignlanguage}[2]{{%
			\expandafter\ifx\csname l@#1\endcsname\relax
			\typeout{** WARNING: IEEEtran.bst: No hyphenation pattern has been}%
			\typeout{** loaded for the language `#1'. Using the pattern for}%
			\typeout{** the default language instead.}%
			\else
			\language=\csname l@#1\endcsname
			\fi
			#2}}
	\providecommand{\BIBdecl}{\relax}
	\BIBdecl
	
	\bibitem{BC}
	T.~M. Cover, ``Broadcast channels,'' \emph{{IEEE} Trans. Inf. Theory}, vol.~18,
	no.~1, pp. 2--14, Jan. 1972.
	
	\bibitem{AWGNBCConverse}
	P.~P. Bergmans, ``A simple converse for broadcast channels with additive white
	{G}aussian noise,'' \emph{{IEEE} Trans. Inf. Theory}, vol.~20, no.~2, pp.
	279--280, Mar. 1974.
	
	\bibitem{MWRCFullExchange}
	L.~Ong, C.~M. Kellett, and S.~J. Johnson, ``On the equal-rate capacity of the
	{AWGN} multiway relay channel,'' \emph{{IEEE} Trans. Inf. Theory}, vol.~58,
	no.~9, pp. 5761--5769, Sept. 2012.
	
	\bibitem{SWoverBC}
	E.~Tuncel, ``Slepian-{W}olf coding over broadcast channels,'' \emph{{IEEE}
		Trans. Inf. Theory}, vol.~52, no.~4, pp. 1469--1482, Apr. 2006.
	
	\bibitem{BCwithSI2UsersOechtering}
	T.~J. Oechtering, C.~Schnurr, I.~Bjelakovic, and H.~Boche, ``Broadcast capacity
	region of two-phase bidirectional relaying,'' \emph{{IEEE} Trans. Inf.
		Theory}, vol.~54, no.~1, pp. 454--458, Jan. 2008.
	
	\bibitem{BCwithSI2UsersKramer}
	G.~Kramer and S.~Shamai, ``Capacity for classes of broadcast channels with
	receiver side information,'' in \emph{Proc. IEEE Inf. Theory Workshop (ITW)},
	Lake Tahoe, CA, Sept. 2007, pp. 313--318.
	
	\bibitem{BCwithSI2UsersGeneral}
	Y.~Wu, ``Broadcasting when receivers know some messages a priori,'' in
	\emph{Proc. IEEE Int. Symp. Inf. Theory (ISIT)}, Nice, France, June 2007, pp.
	1141--1145.
	
	\bibitem{BCwithSI3UsersCommonMessage}
	T.~J. Oechtering, M.~Wigger, and R.~Timo, ``Broadcast capacity regions with
	three receivers and message cognition,'' in \emph{Proc. IEEE Int. Symp. Inf.
		Theory (ISIT)}, Cambridge, MA, July 2012, pp. 388--392.
	
	\bibitem{OechteringG12G22G13G23}
	T.~J. Oechtering, H.~T. Do, and M.~Skoglund, ``Capacity-achieving coding for
	cellular downlink with bidirectional communication,'' in \emph{Proc. Int. ITG
		Conf. Source Channel Coding (SCC)}, Siegen, Germany, Jan. 2010.
	
	\bibitem{OechteringG14G24}
	------, ``Achievable rates for embedded bidirectional relaying in a cellular
	downlink,'' in \emph{Proc. IEEE Int. Conf. Commun. (ICC)}, Cape Town, South
	Africa, May 2010.
	
	\bibitem{BCwithSI3UsersPrivateMessage}
	J.~W. Yoo, T.~Liu, and F.~Xue, ``{G}aussian broadcast channels with receiver
	message side information,'' in \emph{Proc. IEEE Int. Symp. Inf. Theory
		(ISIT)}, Seoul, Korea, June/July 2009, pp. 2472--2476.
	
	\bibitem{Deterministic}
	A.~S. Avestimehr, S.~N. Diggavi, and D.~Tse, ``Wireless network information
	flow: A deterministic approach,'' \emph{{IEEE} Trans. Inf. Theory}, vol.~57,
	no.~4, pp. 1872--1905, Apr. 2011.
	
	\bibitem{Capacity3UsersPrivateMessage}
	B.~Asadi, L.~Ong, and S.~J. Johnson, ``The capacity of three-receiver {AWGN}
	broadcast channels with receiver message side information,'' in \emph{Proc.
		IEEE Int. Symp. Inf. Theory (ISIT)}, Honolulu, HI, June/July 2014, pp.
	2899--2903.
	
	\bibitem{Group4andGroup7}
	------, ``Coding schemes for a class of receiver message side information in
	{AWGN} broadcast channels,'' in \emph{Proc. IEEE Inf. Theory Workshop (ITW)},
	Hobart, Australia, Nov. 2014, pp. 571--575.
	
	\bibitem{Capacity3UsersPrivateMessageParallel}
	J.~Sima and W.~Chen, ``Joint network and {G}elfand-{P}insker coding for
	3-receiver {G}aussian broadcast channels with receiver message side
	information,'' in \emph{Proc. IEEE Int. Symp. Inf. Theory (ISIT)}, Honolulu,
	HI, June/July 2014, pp. 81--85 [Revised Version] Available:
	http://arxiv.org/abs/1407.8409v2.
	
	\bibitem{MultiplexedCoding}
	Z.~Yang and A.~H{\o}st-Madsen, ``Cooperation efficiency in the low power
	regime,'' in \emph{Proc. Asilomar Conf. Signal Syst. Comput.}, Pacific Grove,
	CA, Oct./Nov. 2005, pp. 1742--1746.
	
	\bibitem{IndexCoding}
	Z.~Bar-{Y}ossef, Y.~Birk, T.~S. Jayram, and T.~Kol, ``Index coding with side
	information,'' \emph{{IEEE} Trans. Inf. Theory}, vol.~57, no.~3, pp.
	1479--1494, Mar. 2006.
	
	\bibitem{NetworkCoding}
	R.~Ahlswede, N.~Cai, S.~Li, and R.~W. Yeung, ``Network information flow,''
	\emph{{IEEE} Trans. Inf. Theory}, vol.~46, no.~4, pp. 1204--1216, July 2000.
	
	\bibitem{ITBook}
	T.~M. Cover and J.~A. Thomas, \emph{Elements of Information Theory},
	2nd~ed.\hskip 1em plus 0.5em minus 0.4em\relax Wiley-Interscience, 2006.
	
	\bibitem{DPC}
	M.~H.~M. Costa, ``Writing on dirty paper,'' \emph{{IEEE} Trans. Inf. Theory},
	vol.~29, no.~3, pp. 439--441, May 1983.
	
	\bibitem{MIMOBC}
	H.~Weingarten, Y.~Steinberg, and S.~Shamai, ``The capacity region of the
	{G}aussian multiple-input multiple-output broadcast channel,'' \emph{{IEEE}
		Trans. Inf. Theory}, vol.~52, no.~9, pp. 3936--3964, Sept. 2006.
	
	\bibitem{IndexCodingvsMultiplexingCoding}
	B.~Asadi, L.~Ong, and S.~J. Johnson, ``On index coding in noisy broadcast
	channels with receiver message side information,'' \emph{{IEEE} Commun.
		Lett.}, vol.~18, no.~4, pp. 640--643, Apr. 2014.
	
	\bibitem{NITBook}
	A.~{El G}amal and Y.~H. Kim, \emph{Network Information Theory}.\hskip 1em plus
	0.5em minus 0.4em\relax Cambridge University Press, 2011.
	
\end{thebibliography}

\begin{IEEEbiographynophoto}
{Behzad Asadi}(S'14) received the M.Sc. degree in electrical engineering from the University of Tehran, Iran, in 2008. From 2008 to 2010, he was with WMC Laboratory at the University of Tehran as a research associate, and from 2011 to 2013, he was with ZTE Corporation as a solution manager. Currently, he is pursuing his Ph.D. degree in electrical engineering at the University of Newcastle, Australia. His research interests include information theory, and signal processing in wireless networks.
\end{IEEEbiographynophoto}

\begin{IEEEbiographynophoto}
{Lawrence Ong} (S'05--M'10) received the BEng degree (1st Hons) in
electrical engineering from the National University of Singapore
(NUS), Singapore, in 2001. He subsequently received the MPhil degree
from the University of Cambridge, UK, in 2004 and the PhD degree from
NUS in 2008. He was with MobileOne, Singapore, as a system engineer
from 2001 to 2002. He was a research fellow at NUS, from 2007 to 2008.
From 2008 to 2012, he was a postdoctoral researcher at the University
of Newcastle, Australia. He was awarded a Discovery Early Career
Researcher Award (DECRA) in 2012 and a Future Fellowship in 2014, both
by the Australian Research Council. He is currently a Future Fellow at
the University of Newcastle.
\end{IEEEbiographynophoto}

\begin{IEEEbiographynophoto}
{Sarah J. Johnson} (S'01--M'04) received the B.E. (Hons) degree in
electrical engineering in 2000, and PhD in 2004, both from the
University of Newcastle, Australia. She then held a postdoctoral
position with the Wireless Signal Processing Program, National ICT
Australia before returning to the University of Newcastle where she is
now an Australian Research Council Future Fellow. Sarah's research
interests are in the fields of error correction coding and network
information theory. She is the author of a book on iterative error
correction published by Cambridge University Press.
\end{IEEEbiographynophoto}

\end{document}